\documentclass[12pt,english,preprint]{article}
\usepackage[T1]{fontenc}
\usepackage[latin9]{inputenc}
\usepackage[a4paper]{geometry}
\usepackage{float}
\usepackage{tipa}
\usepackage{amsthm}
\usepackage{amsmath}
\usepackage{amssymb}
\usepackage{graphicx}
\usepackage{setspace}
\makeatletter

\newcommand*\LyXThinSpace{\,\hspace{0pt}}

\makeatother

\usepackage{babel}
\begin{document}

\title{Adaptive gauge method for long-time double-null simulations of spherical
black-hole spacetimes}

\author{Ehud Eilon and Amos Ori}

\maketitle
\noindent \begin{center}
\emph{Department of Physics, }
\par\end{center}

\noindent \begin{center}
\emph{Technion - Israel Institute of Technology, }
\par\end{center}

\noindent \begin{center}
\emph{Haifa 3200003, Israel}
\par\end{center}
\begin{abstract}
\noindent Double-null coordinates are highly useful in numerical simulations
of dynamical spherically-symmetric black holes (BHs). However, they
become problematic in long-time simulations: Along the event horizon,
the truncation error grows exponentially in the outgoing Eddington
null coordinate --- which we denote $v_{e}$ --- and runs out of control
for a sufficiently long interval of $v_{e}$. This problem, if not
properly addressed, would destroy the numerics both inside and outside
the black hole at late times (i.e. large $v_{e}$). In this paper
we explore the origin of this problem, and propose a resolution based
on adaptive gauge for the ingoing null coordinate $u$. This resolves
the problem outside the BH --- and also inside the BH, if the latter
is uncharged. However, in the case of a charged BH, an analogous large-$v_{e}$
numerical problem occurs at the inner horizon. We thus generalize
our adaptive-gauge method in order to overcome the IH problem as well.
This improved adaptive gauge, to which we refer as the \emph{maximal-$\sigma$
gauge}, allows long-$v$ double-null numerical simulation across both
the event horizon and the (outgoing) inner horizon, and up to the
vicinity of the spacelike $r=0$ singularity. We conclude by presenting
a few numerical results deep inside a perturbed charged BH, in the
vicinity of the contracting Cauchy Horizon.
\end{abstract}

\section{Introduction \label{sec:introduction}}

The study of perturbations of the Reissnner-Nordström (RN) background
has been an area of continuing interest in black hole (BH) physics
during the last few decades \cite{1968-Penrose-PRN-Analytical,1973-Simpson-Penrose-PRN-Numerical,1981-Hiscock-PRN-Analytical,1982-Chand-Hartle-PRN-Analytical,1990-Poisson-Israel-PRN-Analytical-mass-function,1991-Ori-PRN-Analytical,1993-Gnedin,1995-Brady-Smith,1997-Burko-Ori Late Time Tails,1997-Burko-Internal-structure,1998-Hod-Piran,2002-Xue-Wang-PRN-Numerical,2013-Reall,2014-Dafermos-PRN-Analytical}.
While the perturbed Kerr geometry is considered to be a more suitable
spacetime model to describe astrophysical black holes, the former
model of a spherical charged BH offers a simpler analysis due to spherical
symmetry. The RN and Kerr spacetimes admit a similar structure of
outer and inner horizons, which implies common behavior in various
phenomena; thus perturbations on RN background were often used as
a ``proto-type'' model for BH structure both in analytical \cite{1968-Penrose-PRN-Analytical,1981-Hiscock-PRN-Analytical,1990-Poisson-Israel-PRN-Analytical-mass-function,1991-Ori-PRN-Analytical,2014-Dafermos-PRN-Analytical}
and numerical \cite{1973-Simpson-Penrose-PRN-Numerical,1993-Gnedin,1995-Brady-Smith,1997-Burko-Ori Late Time Tails,1997-Burko-Internal-structure,1998-Hod-Piran,2002-Xue-Wang-PRN-Numerical,2013-Reall}
analyses.

\begin{figure}[H]
\begin{centering}
\includegraphics[scale=0.6]{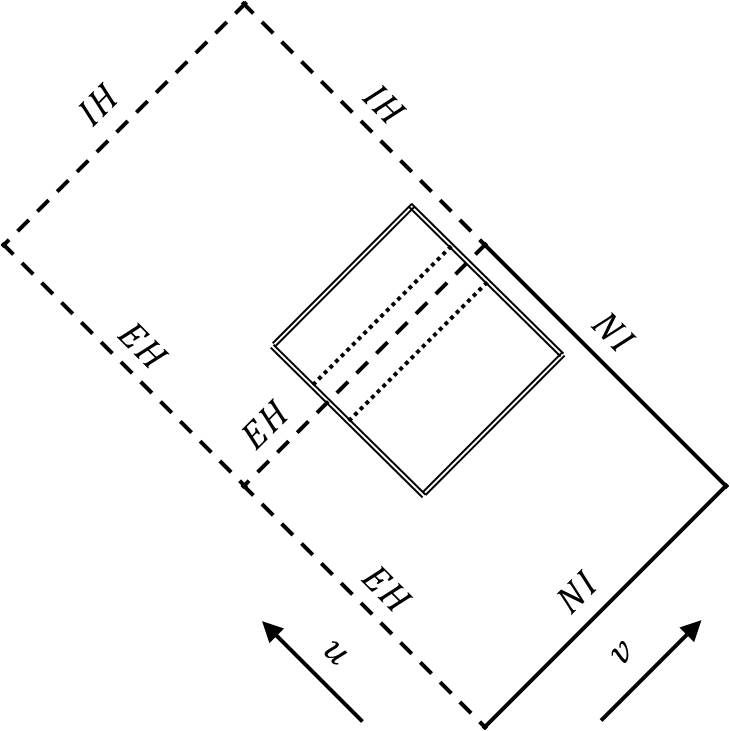}
\par\end{centering}

\raggedleft{}\protect\caption{\label{fig:resprob-RN}Spacetime diagram illustrating the resolution
loss problem in a double-null numerical simulation. The diagram locates
the double-null numerical grid in the RN spacetime diagram; the grid
boundaries are marked by double solid lines. Solid (single) lines
denote null infinity (NI); dashed lines represent the event horizon
(EH) and inner horizon (IH). Dotted lines denote two adjacent grid
lines, the ``outer ray'' and ``inner ray'', at the two sides of
the EH.}
\end{figure}

Double-null coordinates are widely used in numerical analyses of spherically
symmetric spacetimes \cite{Double-Null-Examples1,Double-Null-Examples2,Double-Null-Examples3,Double-Null-Examples4,Double-Null-Examples5,Double-Null-Examples6}.
The choice of double-null coordinates has several advantages: It leads
to relatively simple field equations, it allows convenient coverage
of globally-hyperbolic charts, and simplifies the determination of
the causal structure. However, when it comes to long-time numerical
simulations of (non-extremal) dynamical black-holes, it also has a
serious drawback: unless preventive measures are taken, the numerical
error runs out of control if the integration along the event horizon
(EH) proceeds for a sufficiently long time. The time scale for the
onset of this catastrophic problem, when expressed in terms of the
outgoing null Eddington coordinate $v_{e}$, \footnote{\label{fn:Null_coordinates}We use here $v_{e}$ and $u_{e}$ to denote
the Eddington outgoing and ingoing null coordinates outside the BH.
In particular $u_{e}$ diverges at the EH, whereas $v_{e}$ diverges
at future null infinity. (If the background is dynamical rather than
the static Schwarzschild or RN metric, then the definition of $v_{e}$
and $u_{e}$ becomes somewhat vague, but nevertheless they are chosen
so as to satisfy the appropriate asymptotic properties of the Eddington
coordinates at large $r$ and on approaching the EH). Besides these
specific Eddington coordinates, we also define the \emph{generic}
double-null coordinates $u$ and $v$, with no specific choice of
gauge (see Fig. \ref{fig:resprob-RN}, and also Sec. \ref{sub:Gauge}). } is typically of order $20$-$50$ times the BH mass $M$ (depending
on the numerical parameters, and also on the surface gravity). The
origin of this problem may be traced to dynamics at the horizon's
neighborhood, yet it destroys the numerics in the entire domain of
influence of that neighborhood. This includes the BH interior (at
sufficiently large $v_{e}$), and also the entire strong-field and
weak-field regions at sufficiently large values of the ingoing null
Eddington coordinate $u_{e}$.

To demonstrate this problem, consider two adjacent outgoing null grid
rays, one just before the EH and the next one immediately after the
EH. When simulation time (in $v_{e}$) is long enough, the outer ray
obviously heads towards future null infinity while the inner ray heads
towards either $r=0$ (e.g., in the Schwarzschild metric case) or
Cauchy horizon (in the RN metric case, as illustrated in Fig. \ref{fig:resprob-RN}).
This means that $\delta r$ --- the difference in the area coordinate
$r$ between the two adjacent grid rays at a given $v_{e}$ --- grows
out of control, and the numerics breaks down due to blowing-up truncation
error. In fact, as long as $\delta r$ is $\ll M$, it grows exponentially
with $v_{e}$ --- and so is the truncation error. This exponential
growth of $\delta r$ and truncation error occurs not only in that
specific pair of null grid rays at the two sides of the EH, but in
fact in the entire region near the horizon (both inside and outside),
where $r$ is close to its value at the horizon. 

We have encountered this problem of uncontrolled growth of error when
we attempted to numerically investigate the recently-discovered phenomenon
of outgoing shock wave \cite{2012-Marolf-Ori_Shockwave} that forms
at the outgoing inner horizon of a perturbed charged (or spinning)
BH. Our algorithm for overcoming this problem was developed especially
for this purpose; yet we believe that this numerical algorithm may
be useful for a variety of problems associated with spherical dynamical
BHs. Hence the present paper will be devoted to presenting this algorithm;
the results of our numerical investigation of the shock wave will
be presented elsewhere.

This numerical problem has been previously addressed by Burko and
Ori \cite{1997-Burko-Ori Late Time Tails} using the so-called ``point
splitting'' algorithm, which is based on steady addition of new grid
points at the vicinity of the horizon while propagating in the $v$
direction. This algorithm is very effective, yet it is fairly cumbersome.
Here we propose a different solution to that problem, based on \emph{adaptive
gauge choice} for the ingoing null coordinate $u$. The effect of
this adaptive gauge is similar to crowding the null grid rays in the
vicinity of the horizon. This solution is simpler and more versatile.
It is also generalizable to similar problems that occur near the inner
horizon of a charged black hole. 

The paper is organized as follows: We formulate the physical problem
in terms of the field equations, initial conditions, and coordinate
choice in section \ref{sub:The-Field-Equations}; we present the basic
``four-point'' numerical scheme (for double-null integration of
PDEs in effectively $1+1$ spacetime) in section \ref{sub:The-Numerical-Algorithm}.
Note that sections \ref{sub:The-Field-Equations} and \ref{sub:The-Numerical-Algorithm}
contain fairly standard material, which we nevertheless present here
for the sake of completeness and for establishing our notation. The
EH problem is then described in some detail in section \ref{sec:The-event-horizon-problem};
our solution for this problem, based on the adaptive-gauge method,
is described in section \ref{sec:Horizon-resolving-gauge}. This section
also contains a description of another variant of the algorithm (the
``Eddington-like gauge'' variant), as well as a few examples of
relevant numerical results. The analogous inner-horizon problem is
then discussed in section \ref{sec:The-inner-horizon-problem}, and
our solution to this problem --- an appropriate extension of the adaptive-gauge
method of section \ref{sec:Horizon-resolving-gauge} --- is presented
in section \ref{sec:The-maximal--gauge}. Section \ref{sec:Examples-of-numerical}
contains a few examples of numerical results describing the behavior
of the various fields deep inside the perturbed charged BH at late
time (namely at large $v_{e}$, close to the contracting Cauchy horizon).
We summarize our results and conclusions in section \ref{sec:Discussion}.

\section{The physical system and field Equations \label{sub:The-Field-Equations}}

Let us consider a spherically-symmetric pulse of self-gravitating
scalar field $\Phi$, which propagates inward on the background of
a pre-existing Reissnner-Nordström black hole and perturbs its metric.
The scalar field is neutral, massless, and minimally coupled, and
the initial RN background has mass $M_{0}$ and charge $Q$. We use
double-null coordinate system $(u,v,\theta,\varphi)$, and the line
element is given by:

\begin{equation}
ds^{2}=-e^{\sigma(u,v)}dudv+r(u,v)^{2}d\Omega^{2}\label{eq:line-element}
\end{equation}

\noindent where $d\Omega^{2}=d\theta^{2}+\sin^{2}\theta\,d\varphi^{2}$.
The unknown functions in our problem are thus the scalar field $\Phi(u,v)$
and the metric functions $r(u,v)$ and $\sigma(u,v)$. 

The pulse starts at a certain ingoing null ray $v=v_{1}$; hence,
at $v<v_{1}$ the scalar field vanishes and the metric is just RN
with mass $M_{0}$ and charge $Q$. However, at $v=v_{1}$ the metric
functions are deformed by the energy-momentum contribution of the
scalar field (unless the scalar field perturbation is a test field). 

\begin{center}
\begin{figure}[H]
\begin{centering}
\includegraphics[scale=0.6]{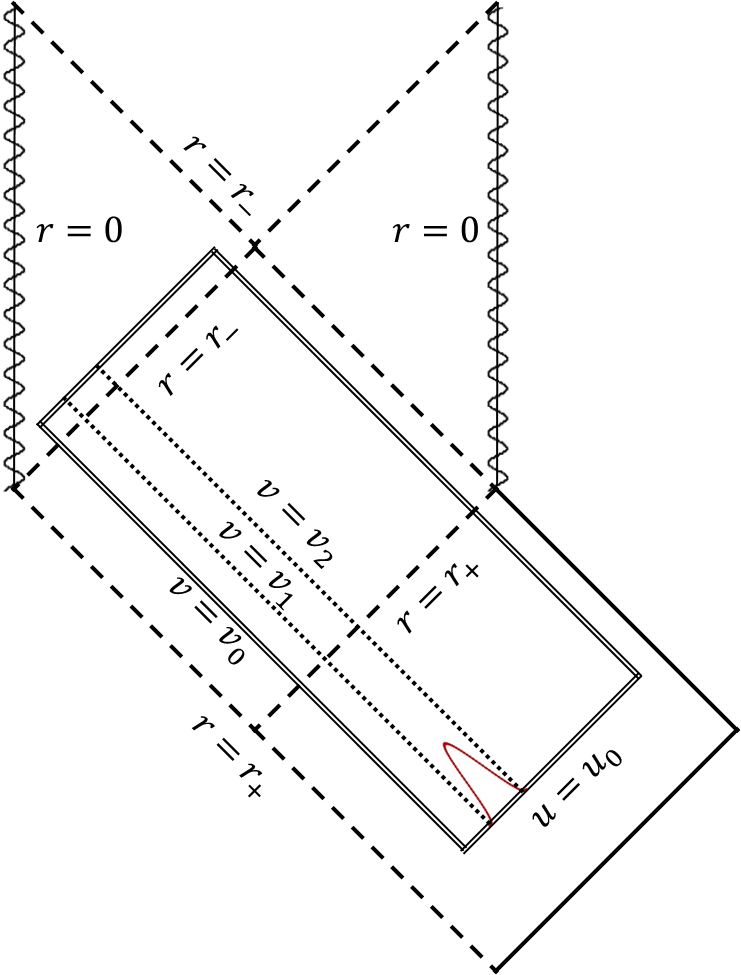}
\par\end{centering}

\raggedleft{}\protect\caption{\label{Fig:Initial-Setup-RN-std}Spacetime diagram illustrating a
typical initial-value setup for the simulation in perturbed RN spacetime.
Grid boundaries are again marked by double lines. Solid lines denote
null infinity; dashed lines represent the event horizon ($r=r_{+}$)
and inner horizon ($r=r_{-}$). Wavy lines denote the time-like $r=0$
singularity of RN. The ingoing scalar field pulse, defined on the
initial ray $u=u_{0}$, is marked by the red curve. It is initially
limited to the range $v_{1}\leq v\leq v_{2}$. In the case of self-gravitating
scalar field the original RN metric is deformed at $v>v_{1}$.}
\end{figure}

\par\end{center}

\subsection{Field equations}

\noindent The field equations and also the energy-momentum contributions
of the scalar and electromagnetic fields are described in some detail
in Appendix \ref{sec: Appendix-A}. In our double-null coordinates
the massless Klein-Gordon equation $\Phi_{;\alpha}^{\,\,;\alpha}=0$
reads
\begin{equation}
\Phi,_{uv}=-\frac{1}{r}(r,_{u}\Phi,_{v}+r,_{v}\Phi,_{u}).\label{eq:  phi_evolutio}
\end{equation}

\noindent The metric evolution is governed by the Einstein equation
$G_{\mu\nu}=8\pi(T_{\mu\nu}^{\Phi}+T_{\mu\nu}^{Q})$, where $T_{\mu\nu}^{\Phi}$
and $T_{\mu\nu}^{Q}$ respectively denote the energy-momentum tensors
of the scalar and electromagnetic fields. Substituting the explicit
expressions (\ref{eq:Energy-Momentum-scalar},\ref{eq:Energy-Momentum-EM})
for $T_{\mu\nu}^{\Phi}$ and $T_{\mu\nu}^{Q}$ in the Einstein equations,
one obtains two evolution equations
\begin{equation}
r,_{uv}=-\frac{r,_{u}r,_{v}}{r}-\frac{e^{\sigma}}{4r}(1-\frac{Q^{2}}{r^{2}}),\label{eq:  r_evolutio}
\end{equation}
\begin{equation}
\sigma,_{uv}=\frac{2r,_{u}r,_{v}}{r^{2}}+\frac{e^{\sigma}}{2r^{2}}(1-\frac{2Q^{2}}{r^{2}})-2\Phi,_{u}\Phi_{,v}\,\,,\label{eq:  sigma_evolutio}
\end{equation}
as well as two constraint equations 
\begin{equation}
r,_{uu}-r,_{u}\sigma,_{u}+r(\Phi,_{u})^{2}=0,\label{eq:  ruu}
\end{equation}
\begin{equation}
r,_{vv}-r,_{v}\sigma,_{v}+r(\Phi,_{v})^{2}=0.\label{eq:  rvv}
\end{equation}

\noindent Overall we have a hyperbolic system of three evolution equations
(\ref{eq:  phi_evolutio}-\ref{eq:  sigma_evolutio}) for our three
unknowns $\Phi(u,v)$, $r(u,v)$, and $\sigma(u,v)$. The two constraint
equations need only be satisfied at the initial hypersurface (see
footnote \ref{fn:const-const}below). \footnote{To avoid confusion we emphasize that $Q$ is a \emph{fixed} parameter
in our model, which is in principle determined by the initial value
of the electromagnetic field, through Eq. (\ref{eq:EM-solution}).
This is in contrast to the mass function $m(u,v)$, which evolves
due to the energy-momentum of the scalar field.}

\subsection{Mass function\label{sub:Mass-function}}

\noindent In addition to our three unknown functions, one may also
be interested in the \emph{mass function}, $m(u,v)$. The mass function
is determined from the metric functions according to \cite{1990-Poisson-Israel-PRN-Analytical-mass-function}

\noindent 
\begin{equation}
g^{\alpha\beta}r,_{\alpha}r,_{\beta}=1-2m/r+Q^{2}/r^{2},
\end{equation}
which in our coordinates translates to

\noindent 
\begin{equation}
m=(1+4e^{-\sigma}r,_{u}r,_{v})r/2+Q^{2}/2r.\label{eq:mass_formula}
\end{equation}

The actual numerical calculation of $m(u,v)$ is based on the numerical
results of $r(u,v)$ and $\sigma(u,v)$ and finite-difference approximation
for the derivatives $r,_{u},r,_{v}$. 

In particular, we shall later refer to the black-hole ``final mass''
($m_{final}$) which we define as the value of the mass function at
the intersection of the event horizon with the final ray of the numerical
grid, $v=v_{max}$.

\subsection{Gauge freedom \label{sub:Gauge}}

The field equations in their double-null form (\ref{eq:  phi_evolutio}-\ref{eq:  rvv})
are invariant under coordinate transformations of the form $v\rightarrow v'(v),u\rightarrow u'(u)$.
In such a transformation, obviously $r$ and $\Phi$ are unchanged,
but $\sigma$ changes according to
\begin{equation}
\sigma\rightarrow\sigma'=\sigma-\ln(\frac{du'}{du})-\ln(\frac{dv'}{dv})\,.\label{eq:  sigma_gauge}
\end{equation}

The choice of optimal gauge may depend on the nature and properties
of the spacetime chart being investigated. For example, if the domain
of integration is restricted to the BH exterior, an Eddington-like
$u$ coordinate may be very convenient. \footnote{\label{fn:Eddington-like}We use here the notion of ``Eddington-like
coordinate'' for null coordinates ($u$ or $v$) which asymptotically
behave like the strict Eddington coordinates $u_{e}$,$v_{e}$ in
RN/Schwarzschild. For example, an Eddington-like $u$ coordinate is
one that diverges on approaching the EH similar to $u_{e}$ in RN/Schwarzschild.
We resort here to Eddington-\emph{like} coordinates (rather than Eddington)
because the spacetime in consideration may be time-dependent, hence
the strictly-Eddington coordinates are not uniquely defined (see also
footnote \ref{fn:Null_coordinates}). } But on the other hand if this domain contains the horizon, then one
has to choose another gauge for $u$, because the Eddington-like $u$
diverges at the horizon. In that case, a compelling $u$ coordinate
is the one that coincides with the affine parameter along the ingoing
initial ray, to which we shall refer as an \emph{affine $u$ coordinate}.
(Recall, however, that in a numerical application, if the domain of
integration has too large extent in $v$, then one would still run
into the aforementioned ``horizon problem''). In all these cases,
an Eddington-like $v$ coordinate may be a reasonable choice, and
the same for affine $v$ coordinate --- regardless of the choice of
$u$.

\subsection{Characteristic initial conditions }

The characteristic initial hypersurface consists of two null rays,
$u=u_{0}$ and $v=v_{0}$ (see Fig. \ref{Fig:Initial-Setup-RN-std}).
\footnote{As can be seen for instance in figures \ref{fig:1_and_3_at_1-RN}
and \ref{fig:1_and_3_at_1-RN-SF}, in our numerical simulations we
actually use the values $u_{0}=v_{0}=0$.} In principle, the initial conditions for the hyperbolic system (\ref{eq:  phi_evolutio}-\ref{eq:  sigma_evolutio})
should consist of the six initial functions $\Phi(u_{0},v)$, $r(u_{0},v)$,
$\sigma(u_{0},v)$ and $\Phi(u,v_{0})$, $r(u,v_{0})$, $\sigma(u,v_{0})$;
however, on each null ray, only two of the three functions may be
freely chosen. The third function should then be dictated by the relevant
constraint equation --- Eq. (\ref{eq:  ruu}) at $v=v_{0}$ and Eq.
(\ref{eq:  rvv}) at $u=u_{0}$. \footnote{\label{fn:const-const}Once the constraint equations were satisfied
by the initial data, these equations are guaranteed to hold in the
future evolution as well (owing to the consistency of the evolution
and constraint equations).} 

In our set-up, we freely choose the two variables $\Phi$ and $\sigma$
along each initial ray, and then determine $r$ along that ray from
the constraint equation. \footnote{\label{fn:residual-gauge-dof}This determination of $r$ is up to
a few free parameters, whose counting is somewhat tricky: Since the
constraint equation is a second-order ODE for $r$, the latter is
only determined (at each initial ray) up to two arbitrary constants,
resulting in four such arbitrary constants overall. One of these constants
is determined by the initial mass at the vertex $(u_{0},v_{0})$,
since the mass function (\ref{eq:mass_formula}) involves $r,_{u}$
and $r,_{v}$. Two other integration constants (one at each ray) are
dictated by the value of $r$ at that vertex. (The initial mass and
initial value of $r$ at the vertex are both freely specifiable.)
This determines three out of the four arbitrary constants. The remaining
constant merely expresses a residual gauge freedom, since $\sigma$
is unaffected by a gauge transformation of the form $v\rightarrow v\,b,u\rightarrow u/b$
for any constant $b>0$. Note that the parameter $Q$ is also determined
by the initial conditions (through the electromagnetic field).} Recall, however, that there is one gauge degree of freedom associated
to each initial ray. Consider for example the ray $u=u_{0}$: The
initial function $\sigma(u_{0},v)$ may be modified in a fairly arbitrary
manner by a gauge transformation $v\rightarrow v'(v)$ (for example
it can always be set to zero by such a transformation). Restated in
other words, the choice of gauge for $v$ can be expressed by the
choice of initial function $\sigma(u_{0},v)$. The same applies to
the other initial ray $v=v_{0}$ and the function $\sigma(u,v_{0})$
defined along it.

We conclude that of the three initial functions that need be defined
along each initial ray, one ($\sigma$) expresses the choice of gauge,
one ($r$) is determined by the constraint equation, and only one
($\Phi$) expresses a true gauge-invariant physical degree of freedom.

\subsubsection{The standard gauge condition\label{sub:The-standard-gauge}}

A compelling gauge choice is the one defined by the initial conditions
\begin{equation}
\sigma(u,v_{0})=0\,,\;\sigma(u_{0},v)=0.\label{eq:standard_gauge}
\end{equation}
We shall refer to it as the ``standard gauge condition''. Note that
this gauge condition implies an affine gauge for both $u$ and $v$.

\section{The basic numerical algorithm \label{sub:The-Numerical-Algorithm}}

We solve the field equations on a discrete numerical $(u,v)$ grid,
with fixed spacings $\Delta u$ and $\Delta v$ in these two directions.
The borders of the grid are defined by four null rays: The two initial
rays $u=u_{0}$, $v=v_{0}$ where initial data are specified, and
the two final rays $u=u_{max}$, $v=v_{max}$. The numerical solution
proceeds along $u=const$ lines in ascending order, from $u=u_{0}$
up to $u=u_{max}$. Along each line the solution is numerically advanced
from $v=v_{0}$ up to $v=v_{max}$. 

The basic building block of our algorithm, the \emph{grid cell}, is
portrayed in Fig. \ref{fig:The-grid-square}. It consists of three
grid points (1,2,3) in which the functions $\Phi$, $r,$ and $\sigma$
are already known (from initial conditions or from previous numerical
calculations), and one point (``4'') in which these functions are
not known yet. The most elementary numerical task is to calculate
the values of these three unknowns --- which we denote by $\Phi_{4}$,
$r_{4},$ $\sigma_{4}$ --- at point 4 . To this end we evaluate the
evolution equations (\ref{eq:  phi_evolutio}-\ref{eq:  sigma_evolutio})
at the central point ``0'' (a fiducial point not included in the
numerical grid), using central finite differences at second-order
accuracy. 

To demonstrate the basic numerical algorithm let us use the generic
symbol $x$ to represent the three unknowns $\Phi$, $r,$ $\sigma$.
The evolution equation for $x$ takes the schematic form 
\begin{equation}
x,_{uv}=F(x,x_{,u},x_{,v})\label{eq: xuv}
\end{equation}
(with some given function $F$). We evaluate this equation at point
0:
\begin{equation}
x,_{uv\,0}=F(x_{0},x_{,u\,0},x_{,v\,0})\equiv F_{0}\,,\label{eq: xuv0}
\end{equation}
where $x_{0}$ stands for $x$ evaluated at point 0 (and similarly
for $x,_{u\,0}$, $x,_{v\,0}$, $x,_{uv\,0}$). For $x,_{uv\,0}$
we have, at second-order accuracy, 

\begin{equation}
x,_{uv\,0}\cong\frac{x_{4}-x_{3}-x_{2}+x_{1}}{\Delta u\Delta v}\,,
\end{equation}
where $x_{i}$ denotes the value of $x$ at point $i$. Next we turn
to evaluate $F_{0}$. For $x_{0}$ we can simply write $x_{0}=(x_{2}+x_{3})/2$.
For $x_{,u\,0}$ and $x_{,v\,0}$, the correct expressions (at second-order
accuracy) would be 
\begin{equation}
x_{,u\,0}\cong\frac{x_{3}-x_{1}+x_{4}-x_{2}}{2\Delta u}\:,\quad\quad x_{,v\,0}\cong\frac{x_{2}-x_{1}+x_{4}-x_{3}}{2\Delta v}\,.\label{eq: xu,xv}
\end{equation}
Substituting these expressions for $x_{0}$, $x,_{u\,0}$, $x,_{v\,0}$
and $x,_{uv\,0}$ in Eq. (\ref{eq: xuv0}) would yield an algebraic
equation (assuming an algebraic function $F$) for the unknown $x_{4}$.
In order to avoid non-linear algebraic equations, we apply a predictor
corrector scheme: 
\begin{itemize}
\item In the first stage (``predictor''), the first-order derivatives
at the R.H.S. of Eq. (\ref{eq: xuv0}) are evaluated at first-order
accuracy only, through the expressions $x_{,u\,0}\approx\frac{x_{3}-x_{1}}{\Delta u}$
and $x_{,v\,0}\approx\frac{x_{2}-x_{1}}{\Delta u}$, hence $F_{0}$
is independent of $x_{4}$. Therefore, at the predictor stage Eq.
(\ref{eq: xuv0}) yields a \emph{linear} equation for $x_{4}$. We
mark the solution of this linear equation by $\tilde{x}_{4}$. This
intermediate value $\tilde{x}_{4}$ is only first-order accurate. 
\item In the second stage (``corrector''), we elevate the expression for
$x_{4}$ from first-order to second-order accuracy. To this end, we
use the above expressions (\ref{eq: xu,xv}) for $x,_{u\,0}$ and
$x,_{v\,0}$ but with $x_{4}$ replaced by $\tilde{x_{4}}$ (a known
quantity). Again, $F_{0}$ is independent of the unknown $x_{4}$,
and we obtain a linear equation for this unknown --- which is now
second-order accurate.
\end{itemize}
When applying this procedure to our actual variables, we find it convenient
to solve the three evolution equations in a specific order: We first
solve Eq. (\ref{eq:  r_evolutio}) to obtain $r_{4}$, then Eq. (\ref{eq:  phi_evolutio})
for $\Phi_{4}$, and finally Eq. (\ref{eq:  sigma_evolutio}) for
$\sigma_{4}$. This way, the predictor-corrector is not needed in
the integration of the last equation (\ref{eq:  sigma_evolutio})
(as can be easily verified from its R.H.S.). We also find it convenient
to choose equal grid spacings, $\Delta u=\Delta v\equiv\frac{M_{0}}{N}$,
where $M_{0}$ is the initial black hole mass and $N$ takes several
values $40,80,160,320,640$ (several values are used on each run to
provide numerical convergence indicators).

This basic numerical scheme usually performs very well --- second
order convergence of $r,\sigma$ and $\Phi$ --- as long as the domain
of integration is not too large (say, a span in $v_{e}$ smaller than
$20$ times $M$). Even if the domain of integration is much larger,
the scheme should still function well if this domain is entirely outside
the BH. But if the domain of integration contains a sufficiently long
portion of the EH, the numerical scheme encounters the aforementioned
``horizon problem'', which we now explore.

\begin{center}
\begin{figure}[H]
\begin{centering}
\includegraphics[scale=0.6]{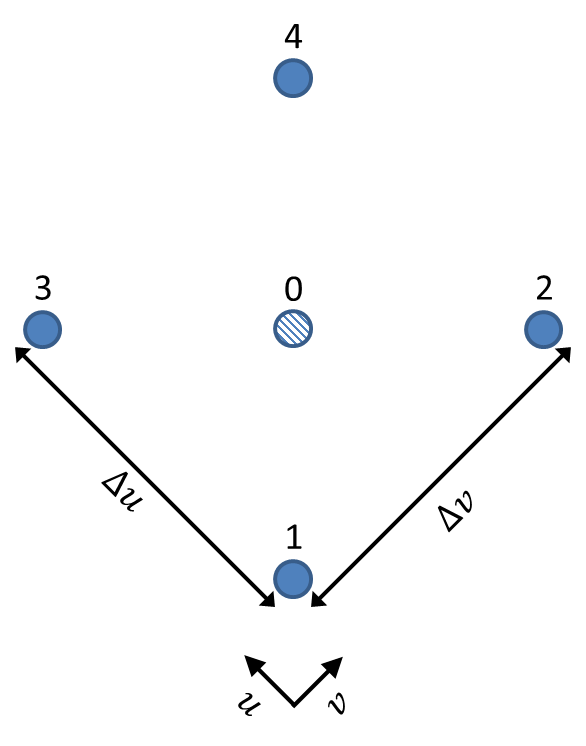}
\par\end{centering}

\protect\caption{\label{fig:The-grid-square}The basic grid cell. Points 1-4 are grid
points; point 0 is an auxiliary point in which the field equations
are evaluated using finite differences. The distances between adjacent
grid points in the $u,v$ directions are $\Delta u$ and $\Delta v$
respectively. }
\end{figure}

\par\end{center}

\section{The event-horizon problem\label{sec:The-event-horizon-problem}}

\subsection{Simplest example: test scalar field \label{sub:Simplest-example:-test}}

To illustrate the EH problem in its simplest form, suppose for example
that we want to numerically analyze the evolution of a test scalar
field $\Phi$ on a \emph{prescribed} RN (or possibly Schwarzschild)
background. Let $U_{k}$, $V_{k}$ be the Kruskal coordinates, and,
recall, $u_{e}$, $v_{e}$ denote the Eddington coordinates (e.g.
of the BH exterior). Suppose that our numerical chart contains (parts
of) the BH exterior as well as interior, so it also contains a portion
of the EH. Then certainly we cannot use the coordinate $u_{e}$ to
cover this chart, because it diverges at the horizon. Instead, we
can in principle use $U_{k}$, or any other $u$-coordinate regular
at the horizon (e.g. the ``affine $u$ coordinate'' introduced in
Sec. \ref{sub:Gauge}) . For the $v$ coordinate the issue of regularity
at the EH does not arise at all, and we can choose $V_{k}$ or $v_{e}$
(or essentially any other $v$ coordinate), it doesn't really matter;
Nevertheless it turns out that the EH-problem is easiest to describe
and quantify in terms of the $v_{e}$ coordinate, so we shall use
this coordinate here. 

Once we choose a horizon-regular $u$ coordinate (and essentially
any $v$ coordinate), both the metric and the scalar field $\Phi$
are perfectly regular across the horizon. This is indeed the situation
from the pure mathematical view-point. However, in a numerical implementation
the situation becomes more problematic. It turns out that along the
horizon's neighborhood the truncation error grows exponentially in
$v_{e}$. In the Schwarzschild case it grows as $\propto e^{v_{e}/4M}$,
and in the more general case as $\propto e^{\kappa_{+}v_{e}}$, where
$\kappa_{+}$ is the surface gravity of the EH. There is a pre-factor
which is proportional to the basic spacing $\Delta u$, but if the
$v_{e}$-extent of the numerical chart is sufficiently large, the
exponentially-growing error eventually gets too large values and the
numerics actually breaks down. 

\begin{center}
\begin{figure}[H]
\begin{centering}
\includegraphics[scale=0.6]{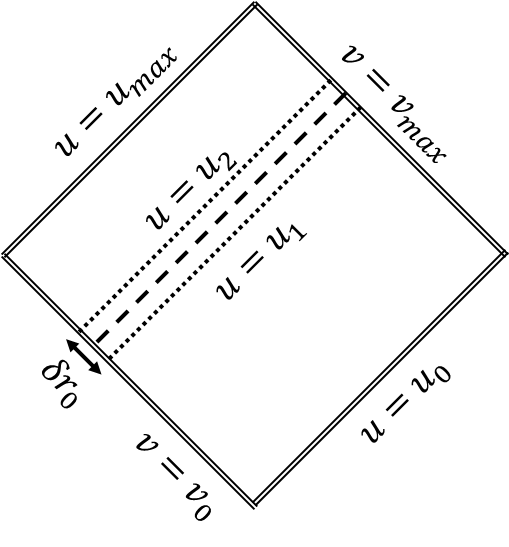}
\par\end{centering}

\raggedleft{}\protect\caption{\label{fig:resprob}A zoom on the numerical grid of Figure \ref{fig:resprob-RN}
illustrating the resolution loss problem. Double solid lines denote
the grid past ($u=u_{0},v=v_{0}$) and future ($u=u_{max},v=v_{max}$)
boundaries. The dashed line is the EH. The dotted lines denote two
adjacent null grid lines close to the event horizon, $u=u_{1}$ and
$u=u_{2}$. The arrow represents the initial radial difference $\delta r_{0}$. }
\end{figure}

\par\end{center}

To understand and quantify this exponential growth of error, consider
two adjacent outgoing null grid lines, denoted $u=u_{1}$ and $u=u_{2}$
(with $u_{2}=u_{1}+\Delta u$): $u_{1}$ is the last $u=const$ grid
line before the EH and $u_{2}$ is the next one, as illustrated in
Fig. \ref{fig:resprob}. At any $v=const$ line, we denote by $\delta r(v)$
the $r$-difference between these two $u$ values: 
\[
\delta r(v)\equiv r(u_{1},v)-r(u_{2},v).
\]
To evaluate $\delta r(v)$ we shall employ the tortoise coordinate
$r_{*}(r)$ which is defined, both inside and outside the EH, as

\begin{equation}
r_{*}=r+\frac{r_{+}^{2}}{r_{+}-r_{-}}\ln|r-r_{+}|-\frac{r_{-}^{2}}{r_{+}-r_{-}}\ln|r-r_{-}|\,.\label{eq:rstar_exact}
\end{equation}
This function is non-invertible, but we can use the near-horizon approximation
for $r(r_{*})$:
\[
r\cong r_{+}\pm C_{\pm}e^{2\kappa_{+}r_{*}},
\]
where the ``$\pm"$ refers to points outside or inside the EH, and
$C_{\pm}>0$ are certain constants whose specific values are unimportant
here. Thus, denoting by $r_{1},r_{2}$ the values of $r$ along the
lines $u=u_{1},u=u_{2}$ respectively (at a given $v$), we write
\[
r_{1}\cong r_{+}+C_{+}\cdot e^{2\kappa_{+}r_{*}},\,\,\,r_{2}\cong r_{+}-C_{-}\cdot e^{2\kappa_{+}r_{*}}.
\]

\noindent Let us now re-express $r_{1,2}$ using Eddington coordinates
$v_{e},u_{e}$, defined (at the two sides of the EH) by $v_{e}=t+r_{*}$
and $u_{e}=\pm(t-r_{*})$, implying $r_{*}=(v_{e}\mp u_{e})/2$. {[}Note
that regardless of the specific gauge that we actually use for $u$
in the numerics, we are allowed to re-express $u$ in terms of $u_{e}$,
which simplifies the analysis here; and the same for $v_{e}$.{]}
We find that 
\begin{equation}
\delta r(v)\equiv r_{1}(v)-r_{2}(v)\cong e^{\kappa_{+}v_{e}}\left[C_{+}e^{-\kappa_{+}u_{e1}}+C_{-}e^{+\kappa_{+}u_{e2}}\right]\,,\label{eq:rstar}
\end{equation}

\noindent where $u_{e1},u_{e2}$ are the Eddington values of $u_{1},u_{2}$.
Notice that the term in squared bracket is a constant. Let us denote
by $\delta r_{0}$ the initial value of $\delta r$ (at $v=v_{0})$.
We find that 
\begin{equation}
\delta r(v)\cong\delta r_{0}\,e^{\kappa_{+}(v_{e}-v_{e0})}\,,\label{eq:  dr(EH)}
\end{equation}
where $v_{e0}$ is the Eddington value corresponding to $v=v_{0}$. 

Since the field equation explicitly depends on $r$ (and its derivatives),
the exponential growth of the difference in $r$ between two adjacent
grid points will inevitably lead to an exponentially growing truncation
error. If the span in $v_{e}$ is large enough, at some point $\delta r(v)$
will become so large (e.g. comparable to $r$ itself) that the finite-difference
numerics will inevitably break down. 

To demonstrate how severe this problem is, suppose for example that
we want to explore the formation of power-law tails of the scalar
field along the horizon of a Schwarzschild (or possibly RN) BH. A
reliable description of tail formation typically requires a $v_{e}$
range of order a few hundreds $M$ (see e.g. Burko and Ori \cite{1997-Burko-Ori Late Time Tails}).
Let us denote this desired $v_{e}$ extent by $\delta v_{e}$, and
pick for example $\delta v_{e}=200M$. Assuming that the exponential
law (\ref{eq:  dr(EH)}) applies, and recalling that in Schwarzschild
$\kappa_{+}=1/(4M)$, we find that $\delta r_{final}\equiv\delta r(v_{max})$
is $e^{50}\,\delta r_{0}\sim5\cdot10^{21}\,\delta r_{0}$. For any
reasonable choice of initial $\Delta u$, in any ``conventional''
(horizon-crossing) gauge for $u$, such $\delta r_{final}$ will be
$\gg M$, rendering the numerical finite-difference scheme useless.
An equivalent assessment can be made in the RN case. However this
phenomenon does not occur in the extremal case, because $\kappa_{+}$
vanishes. 

\begin{figure}[H]
\noindent \begin{centering}
\includegraphics[bb=180bp 350bp 1780bp 940bp,clip,scale=0.3]{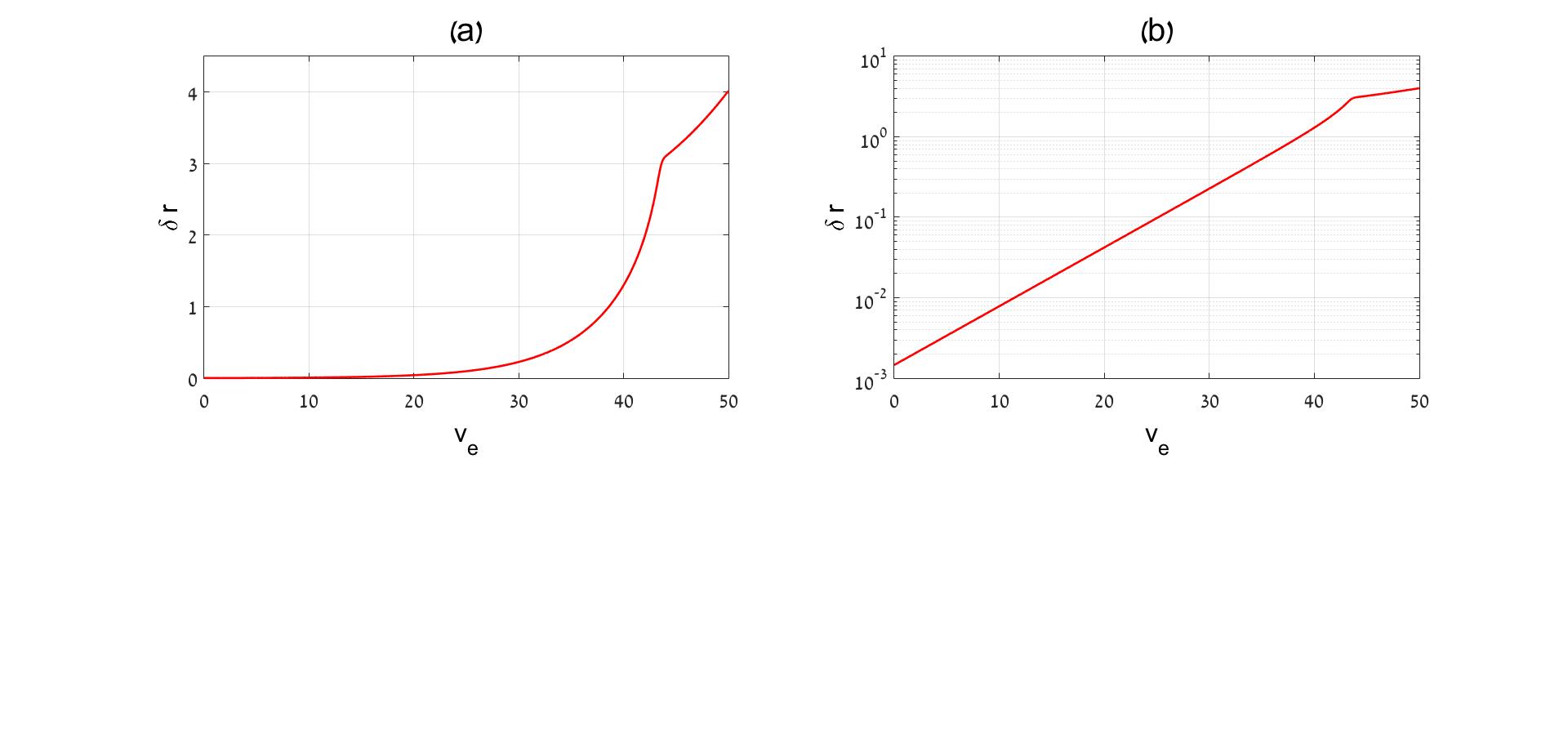}
\par\end{centering}

\protect\caption{\label{fig:Event-Horizon-analytical-Problem} The event horizon problem
as demonstrated from analytical results for the RN spacetime. We consider
here the specific example of mass parameter $M=1.4587$ and charge
parameter $Q=0.95$ in which $Q/M\simeq0.65$, $r_{+}\simeq2.565$,
$r_{-}\simeq0.352$ and $\kappa_{+}\simeq0.168$. The red line represents
the actual $\delta r(v_{e})$ for two adjacent grid lines in different
sides of the horizon, characterized by $r(u_{1},v_{0})=r_{+}+5\times10^{-4}M$
and $r(u_{2},v_{0})=r_{+}-5\times10^{-4}M$. Panel (a) displays $\delta r$
in linear scale, and panel (b) in a logarithmic one. One can see that
$\delta r(v_{e})$ has a long phase of exponential growth, and eventually
it reaches a final value which is larger than $r_{+}$. The breaking
point at $v_{e}\sim44$ represents the point in which the inner line
approaches the vicinity of $r=r_{-}$.}
\end{figure}

\begin{figure}[H]
\noindent \begin{centering}
\includegraphics[bb=150bp 0bp 1800bp 964bp,clip,scale=0.3]{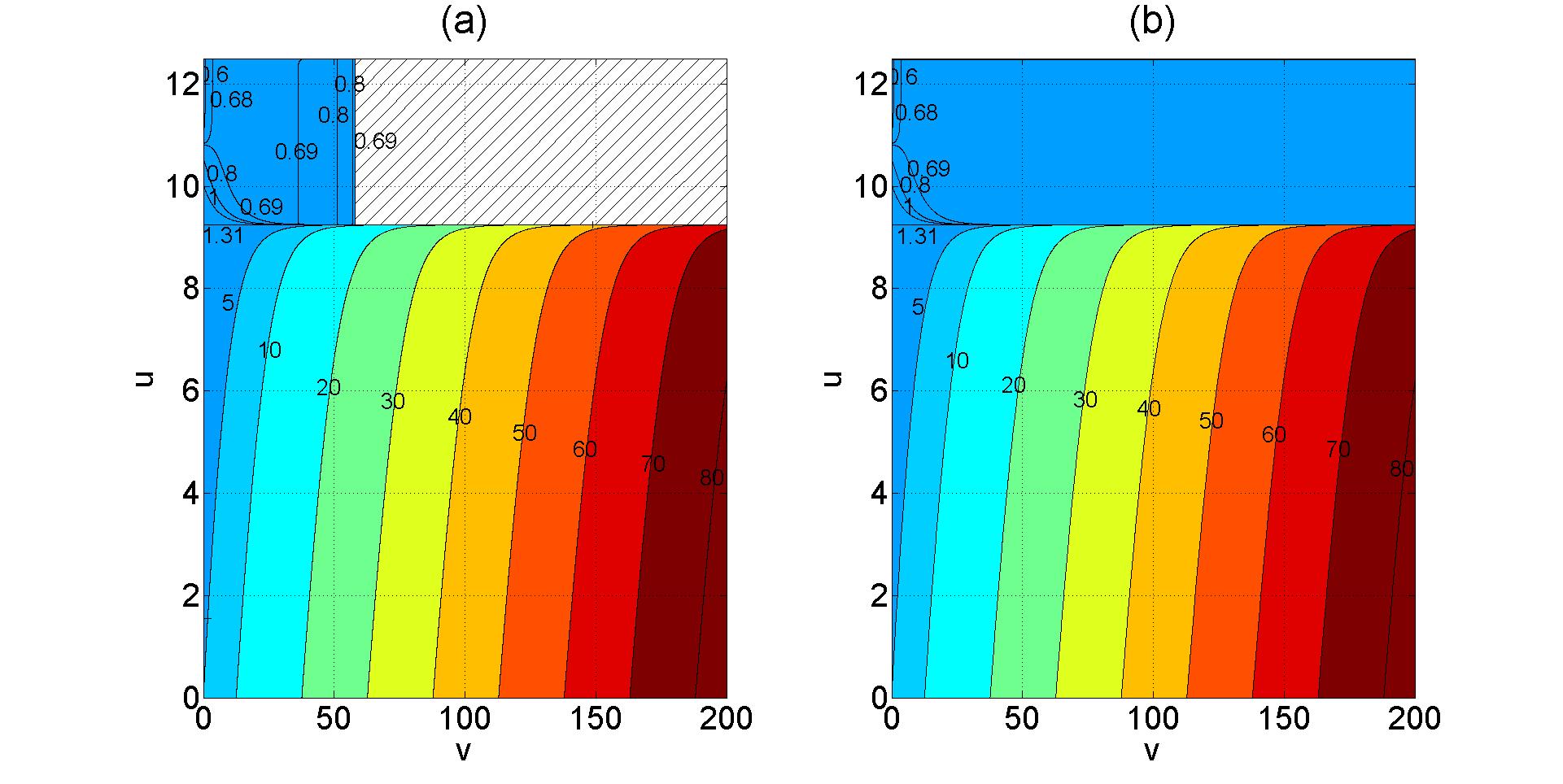}
\par\end{centering}

\protect\caption{\label{fig:1_and_3_at_1-RN} Demonstration of the event-horizon problem
in the case of RN spacetime (with no scalar field). The figure displays
$r(u,v)$ as obtained from the numerical simulation. In this example
the BH mass parameters are $M_{0}=1$ and $Q=0.95$, which yield horizon
radii of $r_{+}\simeq1.312$ and $r_{-}\simeq0.688$. (The initial
value of $r$ is $r(0,0)=5$). Panel (a) displays the numerical results
obtained using the standard gauge condition (\ref{eq:standard_gauge}).
The event-horizon problem expresses itself by the diagonally-striped
patch, the rectangular region at the upper right, where no numerical
data were produced at all. Notice that this region extends also outside
the BH, up to $r\sim60$ in this domain. For comparison, panel (b)
displays $r(u,v)$ for exactly the same initial data, which were numerically
produced using a method designed to resolve the event-horizon problem
(the ``maximal-$\sigma$ gauge'' described in Sec. \ref{sec:The-maximal--gauge}).
The results in panel (b) were converted to the same gauge (\ref{eq:standard_gauge})
as in panel (a) for presentation, in order to allow the comparison.
Both panels are based on numerics with the same resolution $N=640$.
The missing rectangle in panel (a) {[}unlike panel (b){]} demonstrates
the event-horizon problem.}
 
\end{figure}

Note however that the simple exponential growth of $\delta r(v)$
only holds as long as the two outgoing rays are still in the EH neighborhood.
In particular, it is not valid when $\delta r(v)$ is no longer $\ll r_{+}$.
Nevertheless, one can show that whenever Eq. (\ref{eq:  dr(EH)})
predicts (upon extrapolation) values of $\delta r(v_{max})$ that
are comparable or larger than $r_{+}$, the \emph{actual} $\delta r(v)$
between two adjacent rays bordering the EH will always achieve values
of order $r_{+}$ at least (regardless of how small $\delta r_{0}$
was initially). Figure \ref{fig:Event-Horizon-analytical-Problem}
displays $\delta r(v_{e})$ for a RN background with $Q/M\cong0.65$,
and a pair of grid lines (located in both sides of the EH) with initial
separation $\delta r_{0}=10^{-3}M$. The figure was generated using
the analytical relation between $r_{*}$ and $r$ in RN, Eq. (\ref{eq:rstar_exact}). 

So far we restricted the discussion to a pair of adjacent grid lines
$u=u_{1,2}$ at the two sides of the EH. It is straightforward to
show, however, that basically the same phenomenon occurs when the
pair $u_{1},u_{2}$ in consideration is located either outside or
inside the EH (but near the horizon). In all three cases $\delta r$
admits the same exponential growth (\ref{eq:  dr(EH)}). Note that
in the case of two inner grid lines in RN the final separation $\delta r_{final}$
may be fine for large $\delta v_{e}$, because both inner rays runs
to $r=r_{-}$; Yet the numerics still breaks down because in intermediate
$v$ values $\delta r(v)$ becomes $\sim M$. 

\noindent 

So far we considered the numerical solution of the scalar wave equation
on a prescribed RN metric. Let us now address a slightly different
problem: The simulations of the Einstein equations themselves, for
the unknowns $r(u,v)$ and $\sigma(u,v)$, with no self-gravitating
scalar field (and with or without a test scalar field). This numerical
simulation should of course yield the RN geometry, and hence the aforementioned
``horizon problem'', and the consequent numerics break-down, should
occur in this case too. 

This problem is demonstrated in panel (a) in Fig. 6, which displays
the results of the numerical simulation of the Einstein equations
for RN (with no scalar field), using the standard gauge condition
(\ref{eq:standard_gauge}). The BH parameters are $M=1$ and $Q=0.95$.
The figure displays levels of $r$ as a function of $v$ and $u$.
The diagonally-dashed rectangular region represents the domain where
no numerical results were produced at all, due to numerics break-down.
The ``bottom'' boundary of this dashed region is an outgoing $u=const$
ray, which starts at $r\approx r_{+}$ and runs far into the weak-field
region (approaching $r\sim60$ at $v_{max}=200$).\footnote{One can evaluate the equivalent grid side in terms of $v_{e}$. In
the present case, a difference of $\delta v=200$ in the ``standard''
$v$ is equivalent to a difference of $\delta v_{e}\approx172$. In
the case described in the next section (self-gravitating scalar field),
a difference of $\delta v=200$ in the ``standard'' $v$ is equivalent
to a difference of $\delta v_{e}\approx148$. } Notice that this domain of faulty numerics also contains the entire
large-$v$ (say $v>60$) portion of the BH interior.

For comparison, and in order to demonstrate that this ``missing rectangle''
is a mere numerical problem, panel (b) of Fig. 6 displays $r(u,v)$
for exactly the same initial data, presented using the same gauge
for $u,v$ as in panel (a). The only difference is that this time,
for the numerical integration a more sophisticated $u$-gauge was
used (the ``maximal-$\sigma$ gauge'' described in Sec. \ref{sec:The-maximal--gauge}
below, which solves the horizon problem). After numerical integration,
the data were converted back to affine $u$ for presentation, to allow
comparison with sub-figure (a). Thus, as long as the numerics works
properly, the numerical results for $r(u,v)$ in the two panels are
identical. The obvious difference is of course the missing diagonally-dashed
region in the sub-figure (a), which demonstrates the catastrophic
numerical horizon problem. 

\noindent 

\subsection{Self-gravitating scalar field\label{sub:Self-gravitating-scalar-field}}

\begin{figure}[H]
\noindent \begin{centering}
\includegraphics[bb=120bp 0bp 1800bp 964bp,clip,scale=0.3]{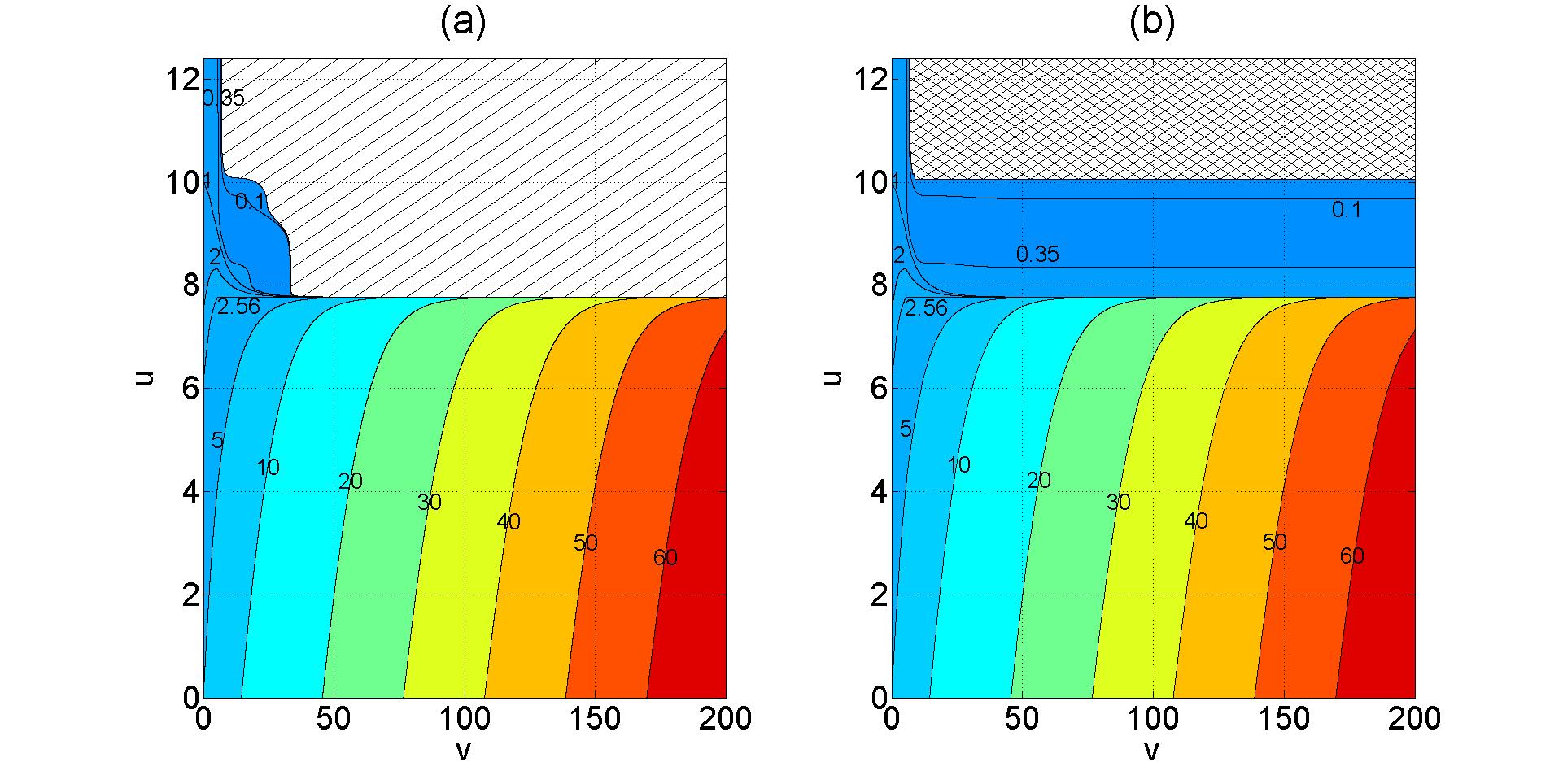}
\par\end{centering}

\protect\caption{\label{fig:1_and_3_at_1-RN-SF} Demonstration of the event-horizon
problem in the case of a spherical charged BH perturbed by self-gravitating
scalar field. Apart from the difference in the physical case, this
figure is similar to figure \ref{fig:1_and_3_at_1-RN}. The BH in
this example has final mass $m_{final}\approx1.4587$ (while $M_{0}=1$)
and charge parameter $Q=0.95$, which yield horizon radii of $r_{+}\simeq2.565$
and $r_{-}\simeq0.352$. The event horizon problem is evident in panel
(a) as a ``hole'' in the grid (the diagonally striped patch). Again,
the problematic domain extends outside the BH to $r\gg M$.  Panel
(b) displays results obtained with the ``maximal-$\sigma$ gauge''
and then converted back to the affine gauge for presentation, to allow
comparison with panel (a). The numerically-covered area in panel (b)
ends around affine $u$ value of $u\sim10$ due to the encounter of
the ray $v=v_{max}=200$ with the close neighborhood of the space-like
$r=0$ singularity; the area in which numerical results are unavailable
due to closeness to the singularity is marked as a criss-cross patch.}
\end{figure}

Of course, the problem described above will also occur with self-gravitating
scalar field, if double-null coordinates are used (with a ``conventional''
gauge for $u$). Due to the mechanism described above, $\delta r(v)$
will grow exponentially in $v_{e}$ along the EH. Since all three
evolution equations (\ref{eq:  phi_evolutio},\ref{eq:  r_evolutio},\ref{eq:  sigma_evolutio})
explicitly depend on $r$, the uncontrolled growth of $\delta r(v)$
will ruin the numerical evolution of all three unknowns. \footnote{In the self-gravitating case the relation $\delta r(v)\propto e^{\kappa_{+}v_{e}}$
still applies along the EH, although it should now be interpreted
as referring to the asymptotic RN metric that evolves at late time.} This catastrophic numerical problem is demonstrated in Fig. \ref{fig:1_and_3_at_1-RN-SF}.
The initial conditions are the same as in Fig. 6 above, the only difference
is that this time there is a self-gravitating scalar field. We use
here (like in Fig. 6) initial mass $M_{0}=1$ and charge $Q=0.95$,
and employ the standard gauge condition (\ref{eq:standard_gauge}).
We send in an ingoing pulse of scalar field, prescribed along the
initial ray $u=u_{0}$ by
\begin{equation}
\Phi(u_{0},v)=\begin{cases}
\begin{array}{c}
A\,\frac{64(v-v_{1})^{3}(v_{2}-v)^{3}}{(v_{2}-v_{1})^{6}}\\
0
\end{array} & \begin{array}{c}
|\,v_{1}\leq v\leq v_{2}\\
|\,\,\,\,\,otherwise
\end{array}\end{cases}\label{eq:pulse}
\end{equation}
with $A=0.115$. This function increases smoothly and monotonically
from zero (at $v=v_{1}$) to a maximum of height $A$ (at $v=\frac{v_{1}+v_{2}}{2}$),
and then decreases smoothly and monotonically back to zero (at $v=v_{2}$).\footnote{We chose this form due to the function properties: it is simple, smooth
and has smooth derivatives up to (and including) second order. It
describes a distinct pulse with well defined beginning, peak and end.
The power $3$ is the minimal power (for this form) in which the function
and the first and second derivatives are smooth. } Throughout this paper, the scalar field vanishes on the initial ray
$u=u_{0}$ -- $\Phi(u,v_{0})=0$. In this case, too, one can see in
panel (a) that the entire large-$v$ portion of the BH interior is
missing (and also a piece of the strong- and weak-field regions outside
the BH). Panel (b) presents the numerical solution with exactly the
same initial data, and displayed using the same gauge for $u,v$ as
in the left panel --- a numerical solution produced using the ``maximal-$\sigma$
gauge'' for $u$ (developed later in Sec. \ref{sec:The-maximal--gauge}).
It again demonstrates that the missing diagonally-dashed region in
panel (a) is a mere numerical artifact. 

One can notice a missing region in panel (b) as well --- the criss-crossed
region at $u>10$. However, this missing region is \emph{not} a numerical
artifact: It results from the fact that the evolution with self-gravitating
scalar field terminates at a spacelike $r=0$ singularity (unlike
in the test-field case).

\noindent 

We point out that if the BH is extremal (or asymptotically extremal,
or even almost extremal) this problem does not arise, because $\kappa_{+}$
vanishes in the extremal case (for numerical example, see \cite{2013-Reall})
. We shall be concerned here with non-extremal BHs. Note also that
if one's only goal was to numerically analyze a test field \emph{outside}
the event horizon of a \emph{prescribed} BH spacetime, then this problem
could be easily circumvented, by just choosing the Eddington coordinate
$u_{e}$. In our case, however, we wish to explore exterior as well
interior regions of the BH; And more crucially, we have a self-gravitating
rather than test field, hence the metric evolves dynamically. In particular,
the location of the EH is not known in advance, which would hamper
attempts to tailor an effectively-Eddington $u$ coordinate to the
EH. 

This EH problem restricted the domain of integration in previous double-null
numerical simulations of dynamical BHs, in spherically-symmetric four-dimensional
models \cite{1993-Gnedin,1995-Brady-Smith,1997-Burko-Ori Late Time Tails,1998-Hod-Piran}
as well as in two-dimensional models \cite{2010-Rama-Pretorious-2D black holes numerics,2011-Ashtekar-Pretorius-2D black holes numerics,2012-Dori Ori evaporation}.
For example, $\delta v_{e}$ was restricted to $\sim20M$ in \cite{1995-Brady-Smith};
In two-dimensional simulations, $\kappa_{+}\,\delta v_{e}$ was restricted
to $\sim20$ in \cite{2011-Ashtekar-Pretorius-2D black holes numerics}
and to $\sim32$ in \cite{2012-Dori Ori evaporation}. 

In some spherically-symmetric simulations of self-gravitating scalar
field, coordinates other than double-null were used for specifying
the components of the metric, but still, the \emph{numerical coordinates}---which
determine the numerical grid points---were double-null. That was the
situation for example in Refs. \cite{1995-Brady-Smith} and \cite{1998-Hod-Piran},
which used a kind of ingoing-Eddington coordinates but with double-null
numerical grid. The problem of exponentially-growing numerical error
along the horizon is encountered in this situation too, for exactly
the same reason: The $r$-difference between two adjacent outgoing
grid lines grows exponentially in $v_{e}$. 

Burko and Ori \cite{1997-Burko-Ori Late Time Tails} developed an
algorithm to address this EH problem, using a variant of adaptive
mesh. More specifically, at certain lines of constant $v$ (separated
from each other by some fixed interval of $v_{e}$), some of the grid
points ``split'' into two: Namely, new grid points are born in the
neighborhood of the EH. This procedure was used in Ref. \cite{1997-Burko-Internal-structure}
for investigating the internal structure of a charged BH perturbed
by self-gravitating scalar field; Similar procedures of adaptive mesh
were applied to the EH problem by other authors \cite{2003-Oren-Piran - Adaptive mesh,2005-Hansen Khokhlov-Novikov - Adaptive mesh,2010-Hong-Yeom et. al - Adaptive Mesh}.
This procedure works very efficiently. However, it is not obvious
if this algorithm would successfully deal with the outgoing shock
wave \cite{2012-Marolf-Ori_Shockwave} that develops at the inner
horizon of a perturbed charged BH. (To the best of our knowledge,
this issue was not tested yet.) Besides, the numerical code implementing
this ``point-splitting'' algorithm is fairly complicated, due to
the appearance of new grid points at certain $v$ values. In particular
it requires interpolations at these new-born points, to determine
the values of the unknown functions there. These interpolations not
only complicate the algorithm but also introduce some random numerical
noise at each such event of ``point splitting''. \footnote{The noisy character of this interpolation-induced error results from
the fact that at a given $v$ value, only some of the points are splitted.} Our goal in this paper is to propose a simpler method for handling
this EH problem, which will not require point-splitting---and which
will successfully resolve the shock at the inner horizon as well.
To this end, in the next sections we shall develop a kind of ``adaptive
gauge'' for the $u$ coordinate, instead of adaptive mesh.

\section{Horizon-resolving gauge \label{sec:Horizon-resolving-gauge}}

In a conventional gauge set-up, one usually picks the function $\sigma_{u}(u)\equiv\sigma(u,v_{0})$
along the initial ray $v=v_{0}$. A particularly simple and convenient
choice is $\sigma_{u}(u)=0$ (the affine $u$ gauge), but any other
choice of this function is allowed. The choice of $\sigma_{u}(u)$
completely fixes the gauge of $u$. \footnote{An analogous procedure may be applied at $u=u_{0}$ to fix the gauge
of $v$.} However, this conventional strategy of gauge choice leads to the
EH problem described above.

To overcome this problem, we use a different strategy for gauge fixing:
We prescribe the function $\sigma(u)$ along the \emph{final} ingoing
ray, $v=v_{max}$. Namely, we define $\sigma_{final}(u)\equiv\sigma(u,v_{max})$.
Choosing different functions $\sigma_{final}(u)$ lead to different
variants of the method. Unless specified otherwise, we shall use here
the simple gauge condition 
\begin{equation}
\sigma_{final}(u)=0,\label{eq:  Final_Gauge}
\end{equation}
to which we shall refer as the ``Kruskal-like'' final gauge. (Another
variant of the method, the ``Eddington-like'' final gauge, is described
below). 

The gauge of $v$ is selected in the more conventional form, via the
initial function $\sigma_{v}(v)\equiv\sigma(u_{0},v)$, as described
in more detail below.

\begin{figure}[H]
\noindent \begin{centering}
\includegraphics[bb=120bp 0bp 1800bp 964bp,clip,scale=0.3]{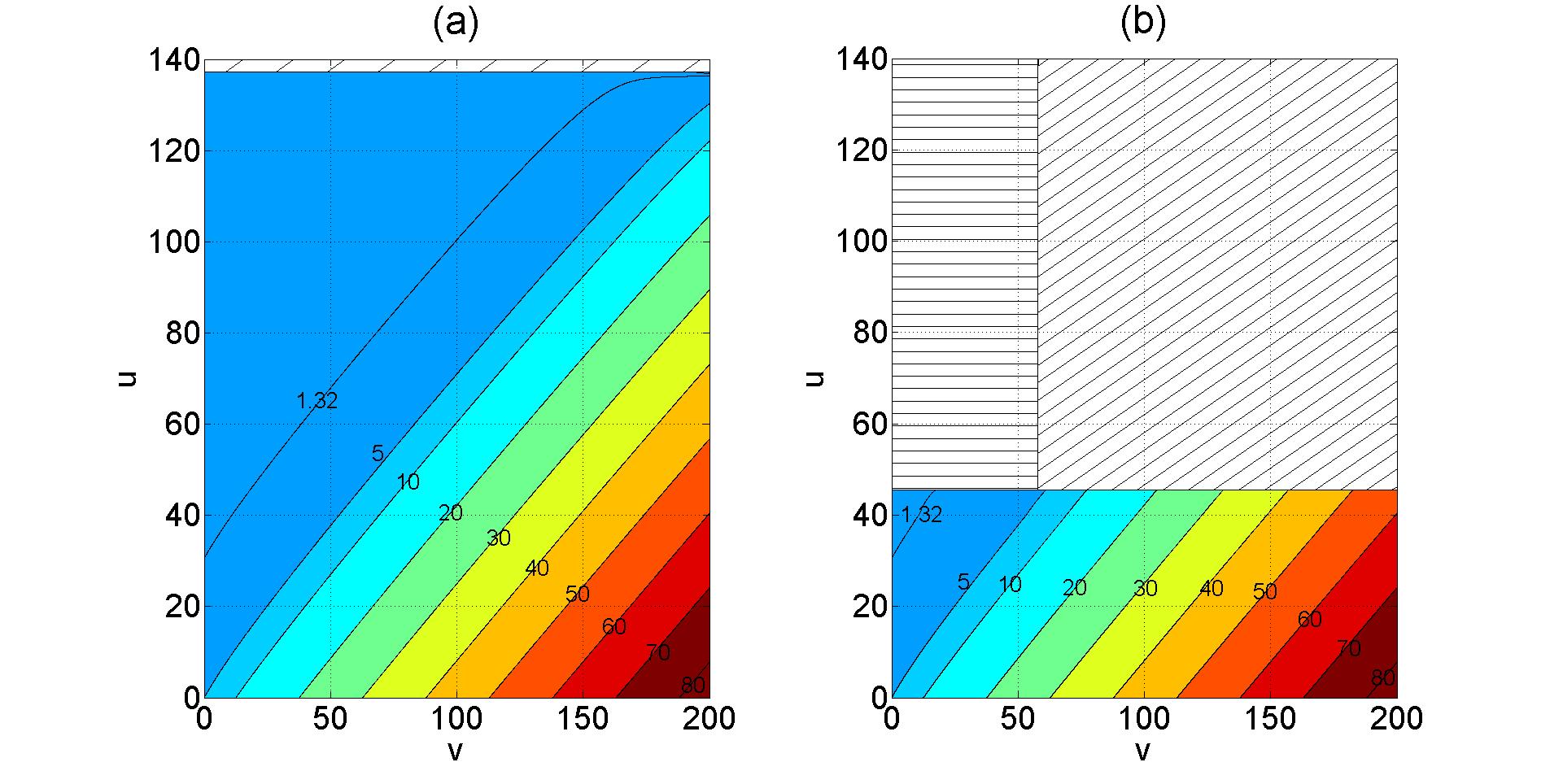}
\par\end{centering}

\protect\caption{\label{fig:1_and_2_at_2-RN} Resolution of the event-horizon problem
in the RN case (no scalar field). Both panels display the numerically
produced function $r(u,v)$ for pure RN initial data. The BH parameters
in this figure are identical to those specified in Fig. \ref{fig:1_and_3_at_1-RN}.
In panel (a) the numerics used the horizon-resolving gauge (\ref{eq:  Final_Gauge}).
In panel (b) the numerics was based on the affine $u$ gauge, although
the results are displayed in the horizon-resolving gauge to allow
the comparison with (a). The diagonally striped patches in both panels
represent domains in which numerical results are unavailable due to
the break-down of the numerics. The figure makes it clear that the
horizon-resolving gauge indeed allows large-$v$ numerical resolution
of the EH (unlike the affine $u$ gauge). In particular, the entire
domain of dependence outside the BH is resolved, as well as a portion
of the BH interior. {[}The narrow diagonally-striped band at the top
of panel (a) indicates the inner-horizon problem, discussed below.
The horizontally striped patch in panel (b) represents a region in
which the affine $u$ gauge functions well but representation of results
in the EH resolving gauge is problematic due to the ``freeze'' of
$r(u,v=0)$.{]} }
 
\end{figure}

\begin{figure}[H]
\noindent \begin{centering}
\includegraphics[bb=120bp 0bp 1800bp 964bp,clip,scale=0.3]{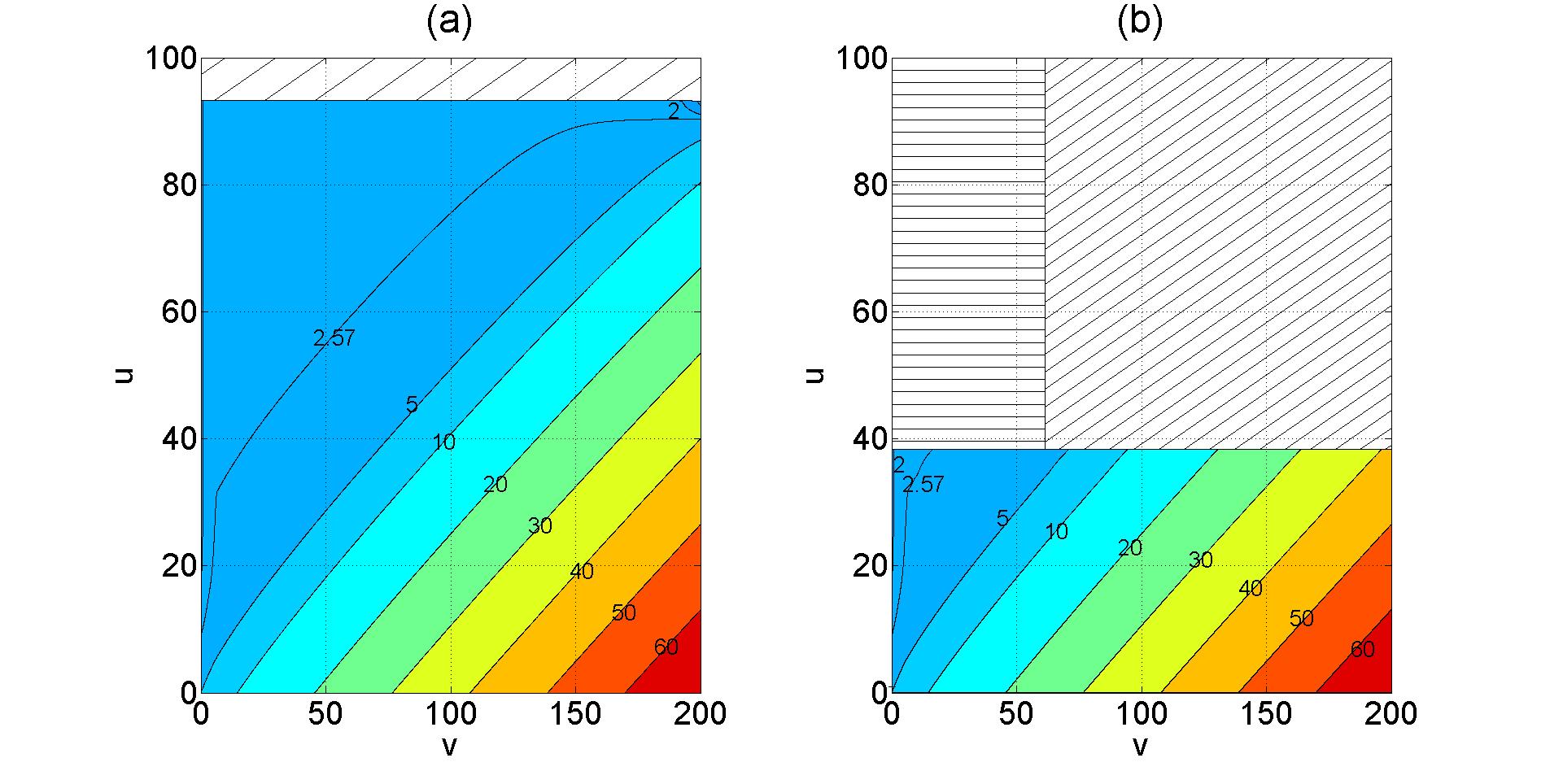}
\par\end{centering}

\protect\caption{\label{fig:1_and_2_at_2-RN-SF} Resolution of the event-horizon problem
in the case of a spherical charged BH perturbed by self-gravitating
scalar field. It is fully analogous to Fig. \ref{fig:1_and_2_at_2-RN},
except the presence of self-gravitating scalar field. The BH parameters
in this figure are identical to those specified in Fig. \ref{fig:1_and_3_at_1-RN-SF}.
In this case too, panel (a) demonstrates the efficient resolution
of the EH problem (but also the occurrence of the IH problem deeper
inside the BH, as indicated by the narrow diagonally-striped band
at the top).}
\end{figure}

The gauge condition (\ref{eq:  Final_Gauge}) resolves the aforementioned
horizon problem. This is demonstrated in Fig. \ref{fig:1_and_2_at_2-RN}
for the RN case, and in Fig. \ref{fig:1_and_2_at_2-RN-SF} for the
self-gravitating scalar field case. In both figures, panel (a) displays
$r(u,v)$ obtained using the horizon-resolving gauge (\ref{eq:  Final_Gauge}).
For comparison, panel (b) displays the numerical solution (with exactly
the same initial conditions) produced by the original affine gauge
and subsequently converted to the horizon-resolving gauge for presentation.
As long as the two numerical simulations function properly, they should
yield the same level curves for $r(u,v)$. Again, the diagonally-dashed
region marks a domain of faulty numerics, where no numerical data
were produced. Panel (b) exhibits the EH problem, the termination
of the numerical solution at the EH, at $u\sim45$. Panel (a) clearly
demonstrates that the horizon-resolving gauge overcomes this problem.
This is the situation in both the RN and scalar-field cases. (Notice,
however, that in both these figures there is a narrow diagonally-dashed
strip at the upper part of panel (a) as well. This strip indicates
a failure of the horizon-resolving gauge (\ref{eq:  Final_Gauge})
somewhere inside the charged BH. We shall return to this issue at
the end of this section.) 

\noindent 

\noindent 

To understand how the EH problem is resolved by the gauge condition
(\ref{eq:  Final_Gauge}), consider again $\delta r(v)$, the ($v$-dependent)
difference in $r$ between two adjacent grid lines $u=u_{1}$ and
$u=u_{2}$ at the horizon's neighborhood. We already saw in the previous
section that near the EH $\delta r(v)\propto e^{\kappa_{+}v_{e}}$.
This now implies that $\delta r(v)$ is everywhere \emph{smaller}
than $\delta r(v_{max})$ (because $v_{e}(v)$ is monotonically increasing).
Thus, as long as $\delta r(v_{max})$ is reasonably small, the horizon
problem is solved. Restated in simple words, in our new gauge set-up
$\delta r(v)$ becomes ``exponentially \emph{small}'' rather than
``exponentially large'' at a typical near-horizon point (e.g. at
the middle of the horizon's section included in the numerical domain). 

To estimate $\delta r(v_{max})$ we express it as $\Delta u$ times
$r_{,u}^{h}$, where $r_{,u}^{h}$ denotes the value of $r_{,u}$
at the intersection point of $v=v_{max}$ and the horizon. The parameter
$\Delta u$ (the numerical spacing parameter) is presumably taken
to be $\ll M$, so it is sufficient to show that $r_{,u}^{h}$ is
of order unity. To this end we consider the constraint equation (\ref{eq:  ruu})
imposed along the ray $v=v_{max}$, which now reads $r,_{uu}=-r(\Phi,_{u})^{2}$.
To simplify the discussion let us assume that at $v=v_{max}$ the
scalar field is so diluted that it has a negligible contribution in
that constraint equation. \footnote{This is a good approximation, recalling the large value of $v_{max}$,
primarily because of  the exponential red-shift along the EH, which
implies exponential decrease of $\Phi,_{u}$.} We thus get $r,_{uu}\approx0$, meaning that $r_{,u}$ is approximately
conserved along the ray $v=v_{max}$. Let $r_{,u}^{0}$ denote the
value of $r_{,u}$ at point ($u=u_{0},v=v_{max}$), namely at the
far edge of the outgoing initial ray. All we need is to make sure
that $r_{,u}^{0}$ is of order unity (or smaller). It turns out that
this parameter depends on the $v$-gauge as well; \footnote{This is so because, since $\sigma=0$ is imposed at that point, a
``stretching'' of $v$ would imply a shrink of $u$. } and in the way we choose our $v$-gauge (by imposing $\sigma(v)=0$
along $u=0$) $r_{,u}^{0}$ indeed turns out to be of order unity.
Hence from the constraint equation $r_{,u}^{h}$ is of order unity
too, which in turn ensures that $\delta r(v_{max})\ll M$, and the
same for $\delta r(v<v_{max})$.

\subsection{Numerical implementation: Initial-value set-up\label{sub:Numerical-implementation:-Initia}}

\subsubsection{Initial data at $v=v_{0}$\label{sub:Initial-data-at-v0}}

From the basic mathematical view-point, the characteristic initial
data required for the hyperbolic system of evolution equations consist
of the values of the three unknowns ($\Phi,r,\sigma$) along the two
initial null rays. We focus here on the data specification along the
ingoing ray $v=v_{0}$ (the initial-data set-up at the other ray $u=u_{0}$
is more conventional, and we shall briefly address it below). We denote
the three initial-value functions at $v=v_{0}$ by $\Phi_{u}(u)$,
$r_{u}(u)$, and $\sigma_{u}(u)$. The first two functions $\Phi_{u}(u)$,
$r_{u}(u)$ can be chosen as usual: Namely, $\Phi_{u}(u)$ can be
chosen at will, and $r_{u}(u)$ should be determined by the constraint
equation (\ref{eq:  ruu}). But the remaining function $\sigma_{u}(u)$
must be chosen in a special manner in order to achieve the desired
gauge condition $\sigma_{final}(u)=0$. We now explain how we choose
the appropriate initial function $\sigma_{u}(u)$. 

We numerically integrate the PDE system along lines of constant $u$,
as described in section \ref{sub:The-Numerical-Algorithm} above.
Suppose that we already integrated the PDE system up to (and including)
a certain grid line $u=const=u_{a}$, and we turn to solve the next
line, $u=u_{a}+\Delta u\equiv u_{b}$. To this end we need to specify
the three initial values for this line, namely $\Phi_{u}(u_{b})$,
$r_{u}(u_{b})$, and $\sigma_{u}(u_{b})$. We set $\Phi_{u}(u_{b})$
by just substituting $u=u_{b}$ in the prescribed initial function
$\Phi_{u}(u)$. Then $r_{u}(u_{b})$ is to be obtained from the constraint
equation (upon integration from $u_{a}$ to $u_{b}$). The more tricky
task is the determination of $\sigma_{u}(u_{b})$. In principle one
can do this by trial and error: For each choice of $\sigma_{u}(u_{b})$,
one can integrate the PDE system along $u=u_{b}$ up to $v_{max}$
and obtain $\sigma_{final}(u_{b})$. Then one can correct $\sigma_{u}(u_{b})$
in an attempt to get a value of $\sigma_{final}(u_{b})$ closer to
the desired value $\sigma_{final}(u_{b})=0$, and so on. This may
be implemented by a simple iterative process, e.g. using Newton-Raphson
method, which converges very rapidly. However, there is another method
which turns out to be much more efficient in terms of computation
time: One can find the desired value of $\sigma_{u}(u_{b})$ by \emph{extrapolation},
based on the values of $\sigma_{u}(u)$ at $u=u_{a}$ and a few additional
preceding $u=const$ lines (namely $u=u_{a}-\Delta u$, $u=u_{a}-2\Delta u$,
... $,u=u_{a}-n\Delta u$). The number of such preceding $\sigma_{u}(u)$
values to be used in the process, would depend on the desired order
of extrapolation. \footnote{\label{fn:In-this-extrapolation}In this extrapolation process one
should of course take into account the ``mismatches'' of $\sigma_{final}$
at the previous $u$-grid points. It therefore follows that the quantity
to be extrapolated (from the previous $u=const$ lines to $u=u_{b}$)
is actually $\sigma_{u}(u)-\sigma_{final}(u)$.} With this procedure, we only need to solve the evolution equations
\emph{once} along any $u=const$ grid line --- just like in the case
of the conventional gauge choice, in which $\sigma_{u}(u)$ is prescribed. 

This procedure works very efficiently. By including a sufficient number
of $u$-grid points in the extrapolation (that is, a sufficiently
large $n$), the mismatch in $\sigma_{final}$ --- namely the deviation
of the latter from zero --- can easily be made third order or even
higher-order in $\Delta u$. In our numerical code we actually use
interpolation order $n=2$. $\sigma_{final}$ is then third-order
in $\Delta u$, and it is typically of order $10^{-8}$ or even smaller.
One should also bear in mind that even this tiny deviation of $\sigma_{final}$
from zero does not necessarily represent an error in the numerical
solution: It merely represents a deviation from the intended gauge
condition $\sigma_{final}=0$ (a deviation which has no effect whatsoever
on the capability of this gauge to achieve its goal of horizon-resolution). 

As already mentioned above, the initial parameter $r_{u}(u_{b})$
is to be determined by integrating the constraint equation (\ref{eq:  ruu})
from $u_{a}$ to $u_{b}$. The discretization of this ODE may involve
also $\sigma_{u}(u_{b})$. But this does not pose any difficulty,
because one can proceed as follows: First obtain $\sigma_{u}(u_{b})$
by extrapolation, and then determine $r_{u}(u_{b})$ from the constraint
equation.

\subsubsection{Initial data at $u=u_{0}$\label{sub:Initial-data-at-u0}}

The initial data along the outgoing ray $u=u_{0}$ are imposed in
a rather conventional manner. We denote the three initial functions
that need be specified along that ray by $\Phi_{v}(v)$, $r_{v}(v)$,
and $\sigma_{v}(v)$. These three functions must satisfy the constraint
equation (\ref{eq:  rvv}). Our view-point concerning the role of
these three initial functions is as described in section \ref{sub:Gauge}:
$\Phi_{v}(v)$ is selected at will, representing the ingoing scalar
pulse; $\sigma_{v}(v)$ may also be selected at will, but it merely
represents one's choice of $v$-gauge; The last function $r_{v}(v)$
is in turn determined by the constraint equation (\ref{eq:  rvv})
--- apart from its initial data, namely the values of $r$ and $r_{,v}$
at the vertex ($u_{0},v_{0}$). 

The choice of $\sigma_{v}(v)$ is subject to one obvious constraint:
It must coincide with the aforementioned final $u$-gauge condition;
that is, $\sigma_{v}(v_{max})=\sigma_{final}(u_{0})$. In our numerical
code we implemented the simplest choice $\sigma_{v}(v)=0$. Consistency
at the vertex is guaranteed, by virtue of our $u$-gauge condition
$\sigma_{final}(u)=0$. 

To summarize, our gauge condition --- throughout the external world
as well as for horizon crossing --- is 
\begin{equation}
\sigma_{final}(u)=0\,,\:\sigma_{v}(v)=0\,.\label{eq:  Final_Gauge:u+v}
\end{equation}

\subsection{The Eddington-like variant\label{sub:The-Eddington-like-variant}}

In some applications one may be interested in time-dependent, dynamically
evolving, BHs, but only in their \emph{external} part, up to the EH
--- and not beyond. This is the situation, for example, when exploring
the influence of the BH dynamics on the scattering of various fields/modes.
A particular example is the issue of how BH dynamics affects the late-time
tails of various modes, e.g. at $r=const$, at future null infinity,
or along the horizon. \cite{1972-Price-Law,2010-Bizon-Power-Law_Tails}

In such applications, there still is the need to numerically resolve
the very neighborhood of the EH, but there is no reason to cross it.
If the metric was prescribed --- e.g. Schwarzschild or RN --- the
Eddington coordinate $u_{e}$ would be optimal for this task. However,
we are concerned here with a-priori unknown dynamical metric. In particular,
we don't know in advance where the EH is. It would thus be helpful
to have at our disposal a gauge-selection method which automatically
yields a $u$ coordinate appropriate for such situations --- namely
an \emph{Eddington-like} $u$ adapted to the dynamically-forming EH.

This goal is easily achieved within the framework of final-gauge condition
for $u$, described above. All that one has to do is to replace the
condition (\ref{eq:  Final_Gauge}) by one appropriate to Eddington-like
$u$. One such example is:
\[
\sigma_{final}(u)=\ln\left(1-\frac{2m_{final}(u)}{r_{final}(u)}+\frac{Q^{2}}{[r_{final}(u)]^{2}}\right)
\]
where $m_{final}(u)$ denotes the mass function evaluated at $v=v_{max}$,
and similarly for $r_{final}(u)$. \footnote{Notice that this expression for $\sigma_{final}(u)$, and similarly
Eq. (\ref{eq:  Final_Gauge-Ed}), are exactly satisfied by $u_{e}$
in Schwarzschild or RN. Hence in the case of a dynamically-forming
spherical BH, imposing any of these conditions yields an asymptotically
Eddington-like $u$ coordinate.} Here, however, we shall use a more convenient and pragmatic gauge
condition which achieves this same goal, but without resorting to
the mass function:

\begin{equation}
\sigma_{final}(u)=\ln[2(r_{,v})_{final}]\label{eq:Final_Gauge-Ed-u}
\end{equation}

\noindent where $(r_{,v})_{final}\equiv r_{,v}(u,v_{max})$. 

In principle, we could apply a similar gauge condition to the initial
ray $u=u_{0}$. Nevertheless, we found it useful (due to certain technical
reasons) to apply the gauge condition:

\begin{equation}
\sigma_{v}(v)=\ln\left(1-\frac{2m_{v}(v)}{r_{v}(v)}+\frac{Q^{2}}{[r_{v}(v)]^{2}}\right)\label{eq:Final_Gauge-Ed-v}
\end{equation}

\noindent which also fits Eddington gauge in RN. However, it is not
trivial that the different gauge conditions (\ref{eq:Final_Gauge-Ed-u},\ref{eq:Final_Gauge-Ed-v})
would yield continuity of $\sigma$ at the corner ($u=u_{0},v=v_{max}$).
Gauge condition (\ref{eq:Final_Gauge-Ed-u}) fixes the $u$ gauge
completely and is unaffected by any change in the $v$ gauge, while
gauge condition (\ref{eq:Final_Gauge-Ed-v}) keeps a residual gauge
freedom (see footnote \ref{fn:residual-gauge-dof}): it is unaffected
by a gauge transformation of the form $v\rightarrow v\,b,u\rightarrow u/b$
(and does not fix either gauge independently). Therefore, we can achieve
continuity at the corner by applying this gauge transformation on
the initial data at $u=u_{0}$ with the appropriate choice of $b$.

To summarize, our gauge condition for the Eddington-like variant of
the horizon-resolving gauge is:

\begin{equation}
\sigma_{final}(u)=\ln[2(r_{,v})_{final}]\,,\,\,\,\sigma_{v}(v)=\ln\left(1-\frac{2m_{v}(v)}{r_{v}(v)}+\frac{Q^{2}}{[r_{v}(v)]^{2}}\right).\label{eq:  Final_Gauge-Ed}
\end{equation}

\subsection{Some numerical results for external tails}

\begin{figure}[H]
\noindent \begin{centering}
\includegraphics[bb=150bp 30bp 1780bp 944bp,clip,scale=0.3]{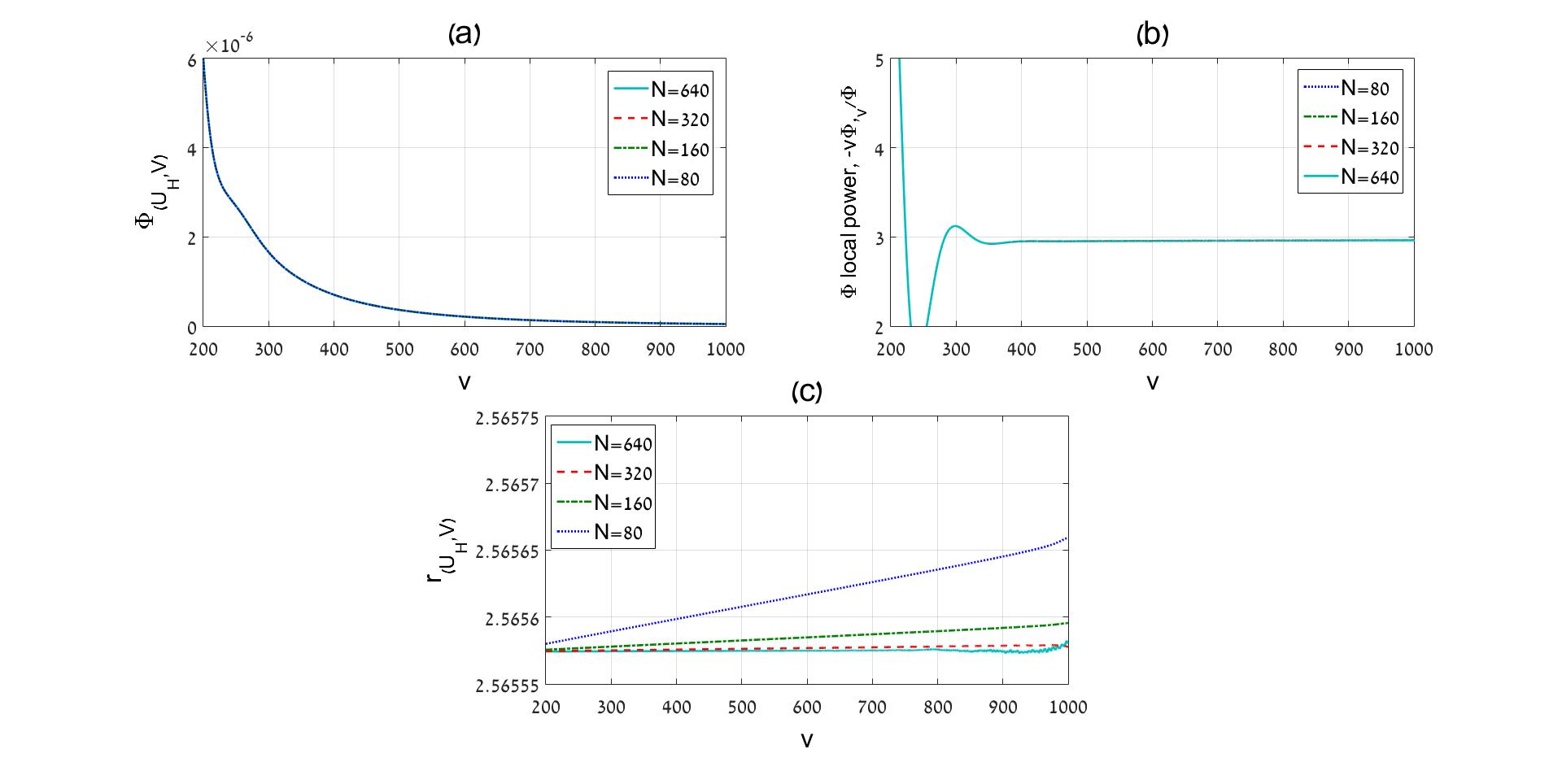}
\par\end{centering}

\protect\caption{\label{fig:Kruskal Results} Numerical results obtained using the
Kruskal-like gauge. The figure describes the late-time behavior of
$\Phi$ and $r$ along the event horizon. Panel (a) displays the decay
of $\Phi$ as a function of $v$. Panel (b) shows the local power
of this decay, defined as $-v\Phi,_{v}/\Phi$.\cite{1997-Burko-Ori Late Time Tails}
In both panels the four resolutions overlap. The asymptotic local
power is approximately $3$, in agreement with Price law\cite{1972-Price-Law}
(see also \cite{2010-Bizon-Power-Law_Tails}). Panel (c) displays
$r$ as a function of $v$. As expected, $r$ is approximately constant
along the curve, with a maximal drift of order $\sim10^{-5}$ along
a range $\Delta v=800$ (for the best resolution $N=640$). This drift
is caused by truncation error.}
\end{figure}

\begin{figure}[H]
\noindent \begin{centering}
\includegraphics[bb=150bp 30bp 1780bp 944bp,clip,scale=0.3]{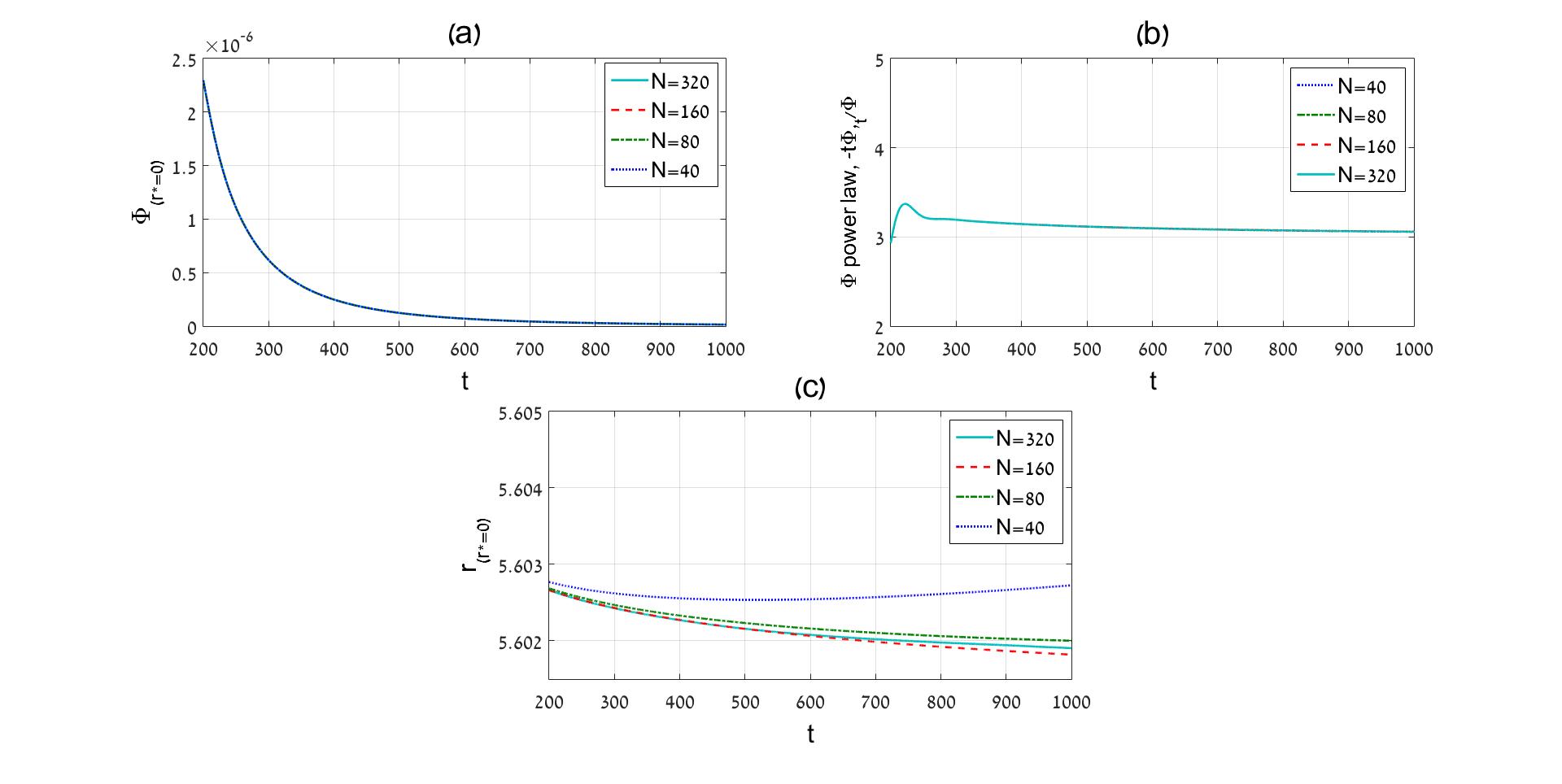}
\par\end{centering}

\protect\caption{\label{fig:Eddington Results} Numerical results obtained using the
Eddington-like gauge. The figure describes the late-time behavior
of $\Phi$ and $r$ along the timelike curve $r*=0$, as functions
of $t$, where $u_{e},v_{e}$ are the numerical Eddington-like coordinates
introduced in section \ref{sub:The-Eddington-like-variant}. Panel
(a) presents the decay of $\Phi$ at large $t$. Panel (b) shows the
local power characterizing this decay. Here again, in both panels
the four resolution overlap, and the asymptotic local power is approximately
$3$, in agreement with Price law. Panel (c) displays $r$ as a function
of $t$ along the same curve $r*=0$. As expected, $r$ is approximately
constant along the curve. However, there is a drift of order $\lesssim10^{-3}$
for the best resolution ($N=320$) along $\Delta t=800$. The evident
drift in the three best resolutions is caused by the fact that the
constructed Eddington-like coordinates, as well as $r*$, fit Eddington
coordinates in pure RN spacetime only asymptotically.}
\end{figure}

We shall refer to the gauge conditions (\ref{eq:  Final_Gauge:u+v})
and (\ref{eq:  Final_Gauge-Ed}) respectively as the \emph{Kruskal-like}
and \emph{Eddington-like} variants of the final $u$-gauge. Since
our main interest is the \emph{interior} of a charged BH, in the next
sections we shall use (and further develop) the Kruskal-like variant,
the gauge condition (\ref{eq:  Final_Gauge:u+v}). Nevertheless, we
bring here some numerical results for both the Kruskal-like (Fig.
\ref{fig:Kruskal Results}) and the Eddington-like (Fig. \ref{fig:Eddington Results})
gauges, concerning late-time tails outside a dynamically-evolving
BH. The initial conditions in these examples are essentially the same
as the ones described in section \ref{sub:Self-gravitating-scalar-field},
namely initial mass $M_{0}=1$ and charge $Q=0.95$, with ingoing
scalar field pulse of the form (\ref{eq:pulse}) with $v_{1}=1$,
$v_{2}=7$, and with $u_{0}=v_{0}=0$. The scalar field amplitude
is again $A=0.115$.\footnote{Due to the small differences between the Kruskal-like and Eddington-like
gauges for $v$ (see section \ref{sub:The-Eddington-like-variant}
above) --- which implies a difference in the physical content of the
pulse $\Phi(u_{0},v)$ --- the final masses in the two runs are slightly
different ($m_{final}\approx1.4587$ in the Kruskal-like case and
$m_{final}\approx1.4595$ in the Eddington-like case); but this tiny
difference in $m_{final}$ is unimportant. } However, in order to reach deep into the late-time regime a larger
$v_{max}$ was used here ($v_{max}=1000$ instead of $200$). 

\noindent 

\noindent 

\subsection{How well does the horizon-resolving gauge penetrate in?}

\begin{figure}[H]
\noindent \begin{centering}
\includegraphics[bb=130bp 0bp 1800bp 944bp,clip,scale=0.3]{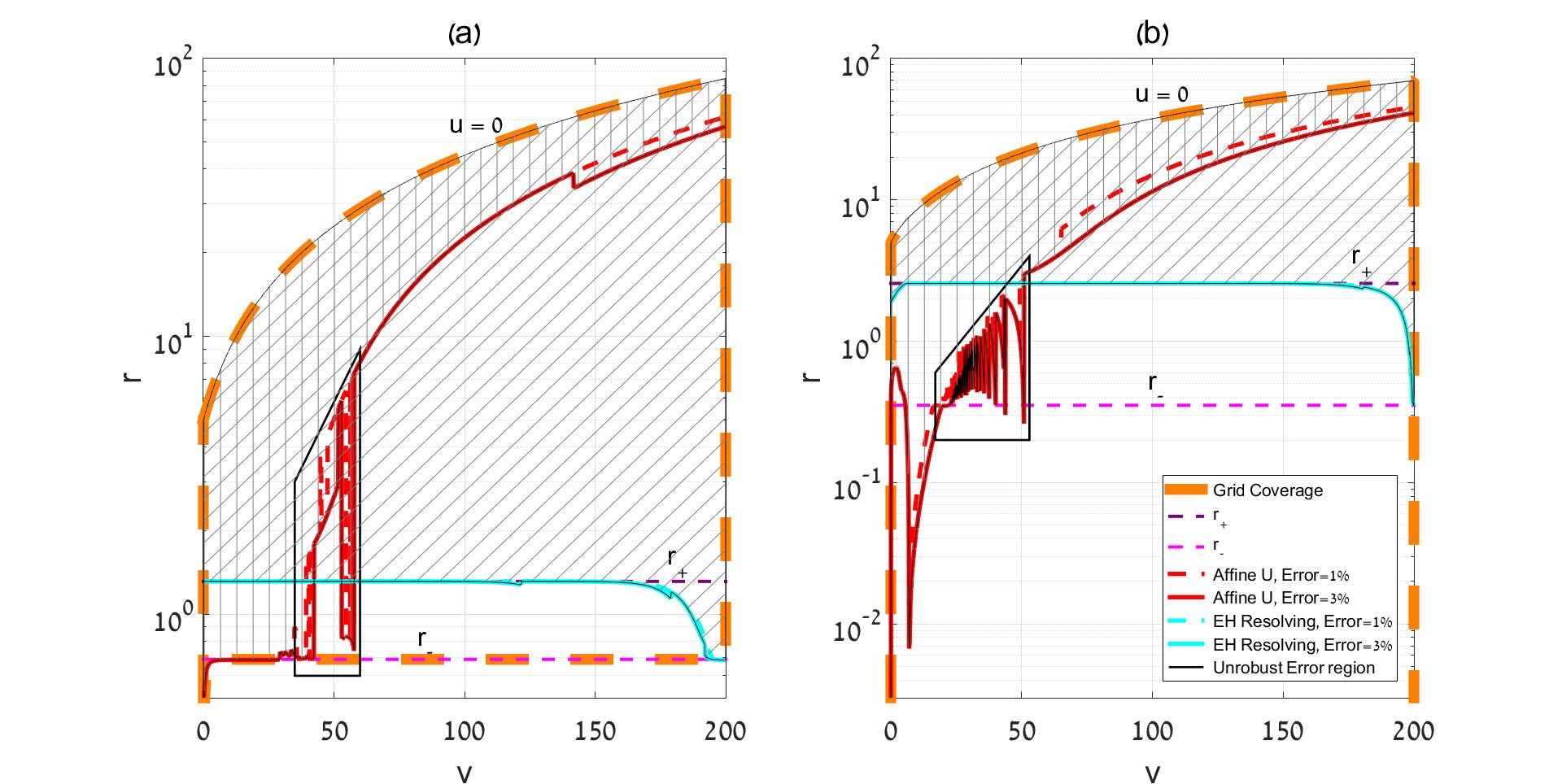}
\par\end{centering}

\protect\caption{\label{fig:LastR-1_and_2} Numerical coverage of the horizon-resolving
gauge in the spacetime of a spherical charged BH. The coverage of
the affine $u$ gauge is also shown for comparison. Panel (a) presents
the case of pure RN initial data, and panel (b) the case of self-gravitating
scalar field. BH parameters of both cases are the same as in figures
\ref{fig:1_and_3_at_1-RN} and \ref{fig:1_and_3_at_1-RN-SF}, respectively.
Both panels presents the last reliable $r$ value on each $v=const$
grid ray as a function of numerical $v$ coordinate (which is common
for both gauges). Numerical reliability is determined by the relative
difference between the two best resolutions $(N=320$ and $N=640$),
and the limit of reliability is chosen as the point in which the relative
difference crosses a threshold $T$, where $T$ is either $1\%$ (dashed
red and cyan curves) or $3\%$ (solid red and cyan curves). The red
curves refer to the affine gauge, and the cyan curves to the horizon-resolving
gauge. The thick dashed orange frame represents the maximal possible
$r$ coverage on each case --- namely boundary of the domain of dependence
of the data. In panel (a), the pure RN case, the bottom orange curve
is computed analytically, it corresponds to the $u=u_{max}$ ray which
starts close to $r=0$ and runs toward $r=r_{-}$ (dashed magenta
line). In panel (b), the case of self-gravitating scalar field, the
bottom orange curve is absent (non-calculable) due to the lack of
analytical solution. The region covered by the affine $u$ gauge is
marked by a vertically striped patch; the region covered by the EH
resolving gauge is marked by diagonally striped patch. The thin black
trapezoid frame in both panels marks a region in which the affine
$u$ gauge has non-robust error behavior. These graphs demonstrate
clearly that the EH resolving gauge has superior numerical coverage
at late times in which it enters the BH (while the affine $u$ gauge
does not even get close to the EH).}
 
\end{figure}

The numerical results presented in the previous subsection made it
clear that the horizon-resolving gauge (like its Eddington-like counterpart)
functions very well outside the BH, all the way up to the EH. But
how deeply can it penetrate into the BH? The narrow diagonally-dashed
strips at the top of the left panels in Figs. \ref{fig:1_and_2_at_2-RN}
and \ref{fig:1_and_2_at_2-RN-SF} indicate numerical problems somewhere
inside the BH. Figure \ref{fig:LastR-1_and_2} provides a more clear
mapping of the regions of good and bad numerics, using a spacetime
diagram with coordinates $r$ and $v$. Panel (a) describes the case
of precisely-RN initial data (i.e. no scalar field), numerically evolved
using either the affine or horizon-resolving gauges for $u$. The
diagonally-dashed region marks the domain of successful numerics using
the horizon-resolving gauge. Similarly, the vertically-dashed region
marks the domain of numerical success of the original affine gauge.
The orange thick dashed frame is the boundary of the maximal possible
domain of integration --- namely the entire domain of dependence of
the characteristic initial data. We primarily focus here on the large-$v$
portion of the diagram (say $v>60$). One can again see that the original
affine gauge does not get even close to the EH (the horizontal dashed
purple line $r=r_{+}$). The horizon-resolving gauge, on the other
hand, covers the entire external region up to the EH. Furthermore,
at very large $v$, close to $v_{max}$ (say $v>180$ in this case)
the numerically-covered region penetrates inside the BH and reaches
$r\sim r_{-}$ at $v_{max}$. However, at smaller $v$ values the
domain of numerical coverage hardly penetrates into $r<r_{+}$. The
situation in the case of self-gravitating scalar field is basically
similar, as seen in panel (b). 

We shall refer to this numerical problem --- this failure of the horizon-resolving
gauge to traverse the IH (or even approach it effectively) --- as
the ``inner-horizon problem''. We shall analyze it in the next section,
and resolve it in Sec. \ref{sec:The-maximal--gauge} by further improving
the gauge.

\noindent 

\section{The inner-horizon problem \label{sec:The-inner-horizon-problem}}

If the black hole has no inner horizon (e.g. an uncharged spherical
BH), the horizon-crossing gauge described above allows efficient numerical
integration of the field equations all the way up to the neighborhood
of the space-like $r=0$ singularity. However, as was just demonstrated,
in the case of a charged BH new difficulties arise while approaching
the inner horizon (IH). This will require us to further generalize
our gauge condition for $u$.

To understand these IH-related difficulties, it would be simplest
to consider again a test scalar field on a fixed RN background. Suppose
that the outgoing initial ray is located outside the BH, and the ingoing
initial ray crosses both the event and inner horizons, as in Fig.
\ref{Fig:Initial-Setup-RN-std}. We also assume that the outgoing
initial ray is long, namely $\delta v_{e}\gg M$. Therefore the domain
of integration contains long sections of both the event and inner
horizons. Now consider two adjacent outgoing null grid lines, $u=u_{1}$
and $u=u_{2}$, this time in the immediate neighborhood of the IH.
Again we consider the quantity $\delta r(v)\equiv r(u_{1},v)-r(u_{2},v)$.
The analysis is analogous to the event-horizon case discussed above,
and we again obtain 
\begin{equation}
\delta r(v)=\delta r_{0}\cdot e^{-\kappa_{-}(v_{e}-v_{e0})}\label{eq:  dr(IH)}
\end{equation}
where $\kappa_{-}>0$ is the inner-horizon surface gravity. Note the
sign difference (in the exponent) between Eqs. (\ref{eq:  dr(IH)})
and (\ref{eq:  dr(EH)}): It actually reflects the difference between
\emph{redshift} (along the EH) and \emph{blue-shift} (along the IH). 

In the ``final $u$-gauge'' described in the previous section, at
$v=v_{max}$ the difference $\delta r(v)$ naturally gets a reasonable
value of order $\Delta u$. \footnote{Considering for example the pure RN case, $\sigma(v_{max})=0$ implies
that $r_{,u}$ is constant along $v=v_{max}$; and as was discussed
in the previous section, this constant is of order unity.} The problem is that this reasonable value of $\delta r$ at $v=v_{max}$
grows exponentially with decreasing $v_{e}$. At $v=v_{0}$ one gets
$\delta r_{0}\approx\Delta u\,e^{\kappa_{-}\delta v_{e}}$. Such an
exponentially-large $\delta r(v)$ naturally leads to a corresponding
huge truncation error, and hence to break-down of the numerics. \footnote{The exponential behavior (\ref{eq:  dr(IH)}) only applies as long
as $\delta r(v)$ is $\ll M$. Nevertheless, this analysis still indicates
that $\delta r(v)$ will fail to be $\ll M$ in some range of $v$. }

In essence, the inner-horizon problem arises from the different signs
of the exponents in Eqs. (\ref{eq:  dr(EH)}) and (\ref{eq:  dr(IH)}):
The positive sign in the former motivated us to use the $\sigma(v_{max})$
``final $u$-gauge'' at the EH, whereas the negative sign in the
latter calls for using ``initial $u$-gauge'' at the IH. 

Naively one might hope that a possible resolution would be to do the
integration in two steps: First, integrate the initial data using
the $u$-gauge $\sigma(v_{max})=0$ up to a certain outgoing line
$u=const\equiv u_{middle}$ located somewhere between the event and
inner horizons, and then integrate the remaining domain $u>u_{middle}$
using the more conventional gauge $\sigma(v_{0})=0$. This procedure
does not work, however: one can show that there is a range of $u$
values, in between the EH and IH, in which neither of the two $u$-gauges
$\sigma(v_{0})=0$ or $\sigma(v_{max})=0$ properly works (because
in both gauges $r_{,u}$, and hence also $\delta r(v)$, gets exponentially-large
values somewhere between $v_{0}$ and $v_{max}$).

To overcome this inner-horizon problem, in the next section we shall
introduce a new $u$-gauge which solves this problem, and allows numerical
resolution of both the event and inner horizons.

\section{The solution: The maximal-$\sigma$ gauge \label{sec:The-maximal--gauge}}

Consider the function $\sigma(v)$ along a line of constant $u$.
We denote by $\sigma_{max}(u)$ the maximal value of this function
in the range $v_{0}\leq v\leq v_{max}$. \footnote{To avoid confusion we emphasize that $\sigma_{max}$ is \emph{not}
defined to be $\sigma(v_{max})$. (Although, these two different objects
may sometimes coincide; in particular, usually they do coincide outside
the BH.) } The value of $v$ at that point of maximal $\sigma$ is denoted $v_{\sigma max}(u)$. 

We shall now impose the gauge condition 
\begin{equation}
\sigma_{max}(u)=0\label{eq:  Maximal_sigma}
\end{equation}
and refer to it as the \emph{maximal-$\sigma$ gauge}. The gauge for
$v$ remains as it was before,

\begin{equation}
\sigma_{v}(v)=0\label{eq:Maximal_sigma_v_gauge}
\end{equation}

Recall that the variable $\sigma$ depends on the gauge choice for
both $v$ and $u$, through Eq. (\ref{eq:  sigma_gauge}). The gauge
of $v$ has already been fixed once and forever by the initial data
at $u=u_{0}$. Along a given $u=const$ line, the effect of changing
the gauge of $u$ would merely be to add $\ln(du/du')=const$ to the
original function $\sigma(v)$. Hence, such a change of $u$-gauge
would not affect the location of the maximum point $v_{\sigma max}$
--- but it would shift $\sigma_{max}(u)$ by that $const$. The specification
of $\sigma_{max}(u)$ thus fixes the $u$-gauge freedom.

With this gauge condition, the inequality 
\begin{equation}
\sigma(u,v)\leq0\label{eq:  Bounded_sigma}
\end{equation}
is satisfied in the entire domain of integration. This boundedness
of $\sigma$ directly reflects on the magnitude of $r_{,u}$, and
hence also on $\delta r$ and the truncation error, as we shortly
explain. This solves the aforementioned numerical difficulties, and
allows efficient long-$v$ numerical integration across both the event
and inner horizons (as well as in the domain beyond the IH).

To understand how the bound on $\sigma$ affects the magnitude of
$r_{,u}$ (and hence $\delta r$ and the truncation error), let us
again consider the simple case of exact RN background. Assume at first
stage that our $v$ coordinate is taken to be the Eddington coordinate
$v_{e}$. Then, if $u$ were taken to be Eddington too, we would have
$e^{\sigma}=2|r_{,u}|$, because in the RN metric in double-null Eddington
coordinates both quantities are equal to $\left|1-2M/r+Q^{2}/r^{2}\right|$.
In a general transformation $u\rightarrow u'(u)$ both $e^{\sigma}$
and $|r_{,u}|$ are multiplied by the same factor $du/du'$, hence
$e^{\sigma}=2|r_{,u}|$ is preserved (as long as $v=v_{e}$). Transforming
next from $v_{e}$ to a general $v$ coordinate, we find that 
\begin{equation}
|r_{,u}|=\frac{1}{2}\frac{\partial v}{\partial v_{e}}\,e^{\sigma}\label{eq:  drdu_Ed}
\end{equation}
is satisfied in a general gauge for $u$ and $v$. Recall, however,
that the ($v$-dependent) quantity $\partial v/\partial v_{e}$ is
determined once and forever by the initial data at $u=u_{0}$; and
in our set-up, this quantity turns out to be of order unity. \footnote{The values in the examples described in sections \ref{sub:Self-gravitating-scalar-field}
and \ref{sub:Simplest-example:-test} are all in the range $0.7\leq\partial v/\partial v_{e}\leq1.5$.} We therefore conclude that in our maximal-$\sigma$ gauge, $|r_{,u}|$
is of order unity at most. Consequently $\delta r$ is bounded by
$\sim\Delta u\ll1$, and the truncation error is well under control. 

So far we considered the case of exact RN background. In the more
general case of nonlinearly-perturbed background metric Eq. (\ref{eq:  drdu_Ed})
no longer holds. Nevertheless, in typical simulations $|r_{,u}|$
and $e^{\sigma}$ are still of same order of magnitude; hence (since
$e^{\sigma}\leq1$) the uncontrolled growth of $\delta r$, and the
related problem of catastrophic truncation error, are prevented. 

It may be instructive to understand the location of the maximal-$\sigma$
curve $v=v_{\sigma max}(u)$ (where in our new gauge $\sigma$ is
set to zero). A detailed analysis of this curve can be done in the
case of RN background. One finds that the integration range $u_{0}\leq u\leq u_{max}$
is divided into three domains: (I) Outside the BH --- and also inside
the BH up to a certain $u$ value which we denote $u_{a}$ --- $v_{\sigma max}(u)=v_{max}$.
(II) Next, in a certain transition domain $u_{a}<u<u_{b}$, $v_{0}<v_{\sigma max}(u)<v_{max}$;
It smoothly moves in this domain from $v_{\sigma max}=v_{max}$ at
$u=u_{a}$ to $v_{\sigma max}=v_{0}$ at $u=u_{b}$. (III) Then, at
$u\geq u_{b}$, we have $v_{\sigma max}(u)=v_{0}$. Stated in other
words, in domain (I) the ``maximal-$\sigma$ gauge'' coincides with
the ``horizon-resolving'' gauge $\sigma(u,v_{max})=0$, in domain
(III) it coincides with the more conventional $u$-gauge $\sigma(u,v_{0})=0$,
and domain (II) is a smooth transition region. 

In the exact RN case --- and with Eddington $v$ --- one finds that
throughout the transition region (II), $v_{\sigma max}(u)$ is precisely
the curve where $r=Q^{2}/M$. Correspondingly, $u_{a}$ is determined
in that case by $r(u_{a},v_{max})=Q^{2}/M$, and $u_{b}$ by $r(u_{b},v_{0})=Q^{2}/M$.
In our actual $v$-gauge (\ref{eq:Maximal_sigma_v_gauge}), $r$ is
still close to $\sim Q^{2}/M$ at $v=v_{\sigma max}(u)$ in region
II. The same applies in the self-gravitating case.

\begin{figure}[H]
\noindent \begin{centering}
\includegraphics[bb=70bp 0bp 1320bp 924bp,clip,scale=0.3]{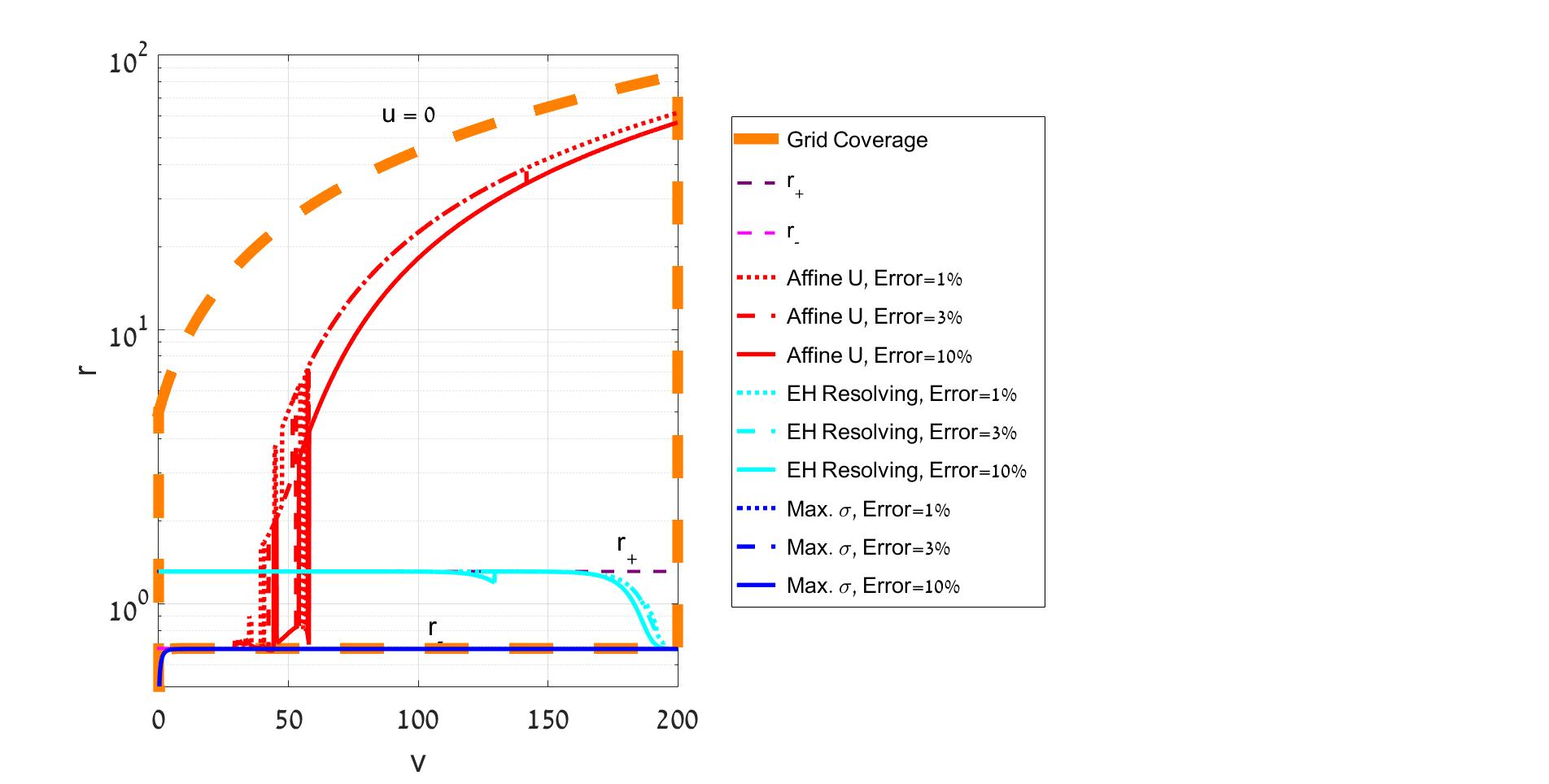}
\par\end{centering}

\protect\caption{\label{fig:LastR-1_2_3-RN} Numerical coverage of the maximal-$\sigma$
gauge in the RN case. The coverage of the affine $u$ gauge and the
horizon-resolving gauge is also shown for comparison. The graph presents
the last reliable $r$ value on each $v=const$ grid ray as a function
of numerical $v$ coordinate and is similar to panel (a) in figure
\ref{fig:LastR-1_and_2}, except that (i) here it shows coverage differences
between \emph{three} gauges (the affine $u$ gauge in red, the EH
resolving gauge in cyan, and the maximal-$\sigma$ gauge in blue),
and (ii) it represents each gauge with \emph{three} sets of lines,
each one for a different relative-error threshold $T$ (dotted lines
- $1\%$; dashed lines - $3\%$; and solid lines - $10\%$). Again,
the thick dashed orange frame represents the maximal possible $r$
coverage in this case. The graph demonstrates clearly that the maximal-$\sigma$
gauge has superior numerical coverage: Note that all three blue curves
(representing the three error levels in the maximal-$\sigma$ gauge)
overlap in this diagram --- and they also overlap with the bottom
orange curve. This indicates that the maximal-$\sigma$ gauge essentially
covers the entire domain of dependence.}
 
\end{figure}

\begin{figure}[H]
\noindent \begin{centering}
\includegraphics[bb=130bp 0bp 1800bp 914bp,clip,scale=0.3]{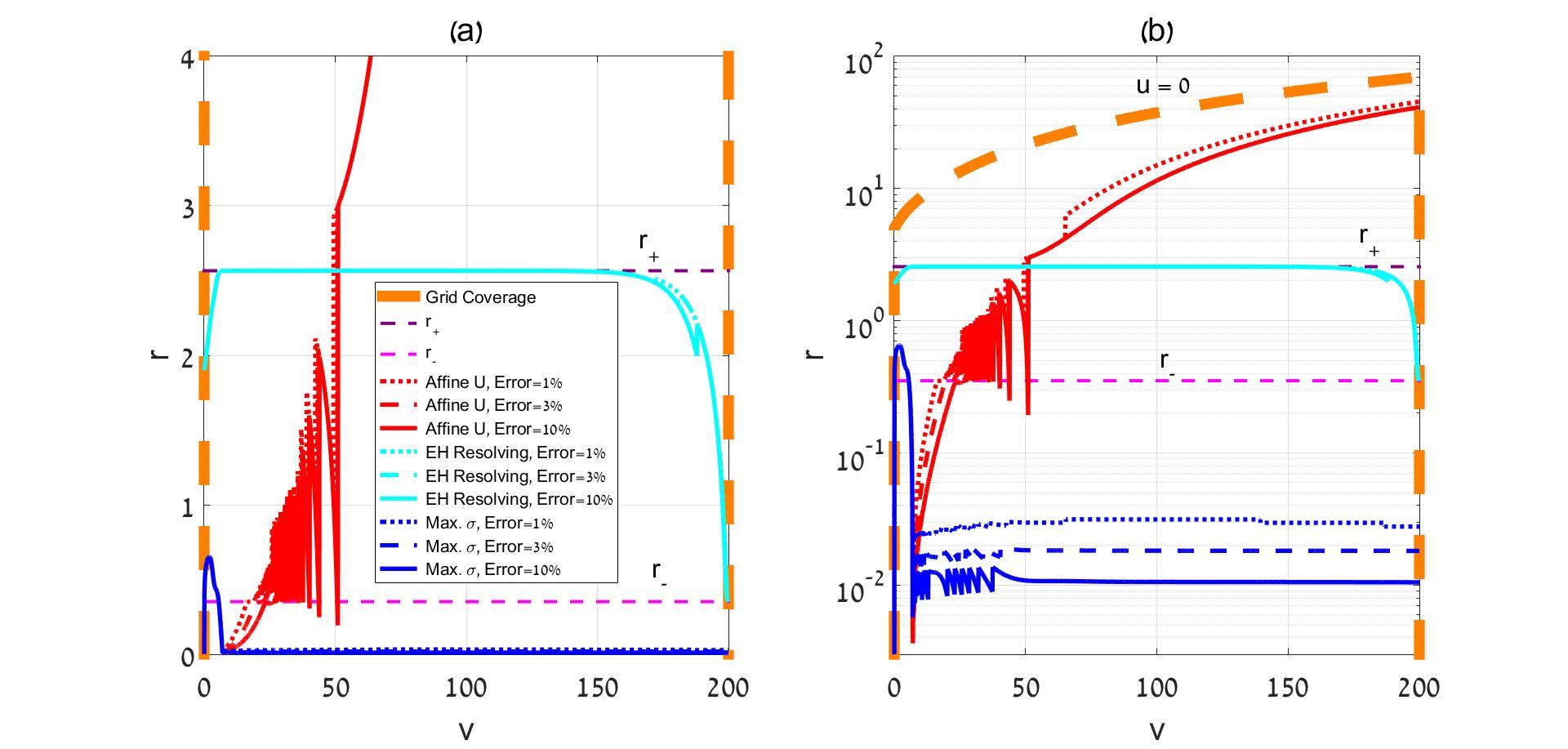}
\par\end{centering}

\protect\caption{\label{fig:LastR-1_2_3-RN-SF} Numerical coverage of the maximal-$\sigma$
gauge in the presence of self-gravitating scalar field. The coverage
of the affine $u$ gauge and the horizon-resolving gauge is also shown
for comparison. The meaning of the various curves and symbols in this
figure is the same as in Fig. \ref{fig:LastR-1_2_3-RN}, the only
difference is the presence of the scalar field. This figure uses both
a linear scale (a) and logarithmic scale (b) for the $r$ axis. The
blue curve in panel (a) approximately overlaps with the $r=0$ line,
indicating that the maximal-$\sigma$ gauge provides efficient numerical
coverage up to very close to the spacelike $r=0$ singularity. (The
significant deviation of the blue curve from $r=0$ at $v\lesssim10$
does not represent a numerical problem: It is a true feature of the
function $r(u_{max},v)$, which merely reflects the spacelike character
of the $r=0$ singularity.) From the logarithmic scale in panel (b)
one can see that the numerical coverage extends up to $r$ values
of order a few times $10^{-2}$ (depending on the required accuracy).}
\end{figure}

This new gauge (\ref{eq:  Maximal_sigma},\ref{eq:Maximal_sigma_v_gauge})
provides a robust solution to the aforementioned EH and IH problems.
This is demonstrated in Figs. \ref{fig:LastR-1_2_3-RN} and \ref{fig:LastR-1_2_3-RN-SF},
which correspond to the RN and self-gravitating field cases respectively.
Both figures display the domain of numerical validity of the maximal-$\sigma$
gauge on a spacetime diagram in ($r,v$) coordinates. For comparison,
the figures also display the domain of numerical validity of the two
former gauges, namely the affine gauge and the ``horizon-resolving''
gauge. In Fig. \ref{fig:LastR-1_2_3-RN} (Pure RN initial data), one
sees that the maximal-$\sigma$ gauge effectively resolves the entire
domain of dependence of the characteristic initial data (the dashed
orange frame). In the more interesting case of self-gravitating scalar
field the situation is more subtle: In this case, the ``bottom''
boundary of the domain of dependence is very close to the spacelike
singularity at $r=0$. The singular nature of this boundary makes
it very hard for the numerics to approach it. Nevertheless, the maximal-$\sigma$
gauge numerics allows reliable coverage up to $r$ values of order
few times $10^{-2}$, as can be seen in panel (b) of Fig. \ref{fig:LastR-1_2_3-RN-SF}. 

\noindent 

\noindent 

\subsection{Numerical implementation: Initial-value set-up}

Our new gauge condition is given in Eqs. (\ref{eq:  Maximal_sigma},\ref{eq:Maximal_sigma_v_gauge}).
The procedure of initial-value setup along $v=v_{0}$ is almost the
same as the one described in section \ref{sub:Numerical-implementation:-Initia}.
There is one obvious difference in the setup of the initial value
for $\sigma_{u}(u)$: Instead of extrapolating the quantity $\sigma_{u}(u)-\sigma_{final}(u)$
(see footnote \ref{fn:In-this-extrapolation}), we now extrapolate
$\sigma_{u}(u)-\sigma_{max}(u)$. This reflects the fact that we are
now attempting to nullify $\sigma_{max}(u)$ rather than $\sigma_{final}(u)$.
The initial-value setup along the ray $u=u_{0}$ is identical to the
one described in section \ref{sub:Numerical-implementation:-Initia}. 

There remains the technical issue of evaluating $\sigma_{max}(u)$,
the maximal value of $\sigma$ along each $u=const$ line (required
for the extrapolation process that determines $\sigma_{u}(u)$.) The
first obvious step is to find the grid point of maximal $\sigma$
along the $u=const$ line under consideration. However, this is not
always sufficient. As was mentioned above, the $u_{0}\leq u\leq u_{max}$
range is divided into three domains, according to the value of $v_{\sigma max}(u)$.
In domain I we have $v_{\sigma max}(u)=v_{max}$, hence $\sigma_{max}(u)$
is simply the value of $\sigma$ at the last grid point ($v=v_{max}$).
Similarly in domain III, $\sigma_{max}(u)$ is the value of $\sigma$
at the first grid point ($v=v_{0}$). However, in the transition domain
II the situation is slightly more complicated: In a generic $u=const$
line in this domain, $v_{\sigma max}(u)$ will be located in between
two grid points. If one approximates $\sigma_{max}(u)$ by one of
these two points (e.g. the one with the largest $\sigma$), this leads
to a discretization noise (amplified by the extrapolation process)
which may seriously damage the accuracy of our second order extrapolation.
Therefore we need here a more precise evaluation of $\sigma_{max}(u)$.
To this end, we carry a second-order interpolation of $\sigma(v)$
(along the $u=const$ line) in the neighborhood of the grid point
of maximal $\sigma$, and determine $\sigma_{max}(u)$ from this interpolation.
This procedure eliminates most of the numerical noise in the extrapolation.

\section{Some numerical results deep inside the charged black hole\label{sec:Examples-of-numerical}}

The maximal-$\sigma$ gauge described above enables us to efficiently
prob the innermost zone $0<r\leq r_{-}$ of the perturbed charged
BH, even at very late times (i.e. large $v_{e}$). As an example,
we choose here to explore the behavior of $r$ and $\Phi$ along the
contracting Cauchy horizon (CH) and its neighborhood, in the case
of self-gravitating scalar field. All results below correspond to
the same numerical simulation of a charged BH with a self-gravitating
scalar field, with the same set of parameters as described in Sec.
\ref{sub:Self-gravitating-scalar-field}. Note that in all figures
throughout this section, the numerical results were produced with
two different resolutions, $N=320$ (dashed curves) and $N=640$ (solid
curves), in order to control the numerical error. In almost all cases
the dashed curve is visually undistinguished from the solid curve,
indicating negligible numerical error. (The exceptions are panels
(c) and (d) in Fig. \ref{fig:Phi_of_r}, where the high zoom level
in $r$ ($\lesssim10^{-5}$) allows us to see the small differences
between the resolutions). 

We argue that the behavior of various quantities along the CH should
be well represented by the lines $v=const$ with sufficiently large
value of the ``$const$''. To theoretically examine this issue let
us consider the case of RN background with test scalar field. Then
two separate issues are involved here: (i) In the domain beyond the
(outgoing section of the) IH, all outgoing null rays run to $r=r_{-}$
at the limit $v_{e}\to\infty$. The ``distance'' from the IH (in
terms of $r$ values) at finite $v_{e}$ scales as $e^{-\kappa_{-}v_{e}}$.
(ii) Wave-scattering dynamics takes the asymptotic form $\Phi(u,v)\cong f_{u}(u)+f_{v}(v)$
near the IH, and at sufficiently large $v$ (namely $v_{e}\gg m$)
we have $f(v)\propto(v_{e}/m)^{-3}$. The scalar field thus approaches
the CH limiting function $\Phi\simeq f_{u}(u)$ at $v_{e}\gg m$.
With $v_{max}=200$ the first issue (i) is perfectly well addressed
as $e^{-\kappa_{-}\delta v_{e}}$ is extremely small. Concerning the
other issue (ii), at $v=200$ wave dynamics is considerably suppressed
--- namely $f_{v}(v)$ is negligible to a fairly good approximation
(although here the approximation is not as good as it is in regards
to the first issue, because the decay here is inverse-power rather
than exponential). 

In the presence of self-gravitating scalar field the near-CH dynamics
becomes more complicated and we shall not discuss it here. Nevertheless,
the $v=const$ lines with sufficiently large ``$const$'' should
still mimick the CH. The quality of this ``CH-mimicking'' approximation
for large-$v$ rays will be verified by the numerical results that
we shortly display. 

\begin{figure}[H]
\noindent \begin{centering}
\includegraphics[bb=160bp 0bp 1000bp 900bp,clip,scale=0.3]{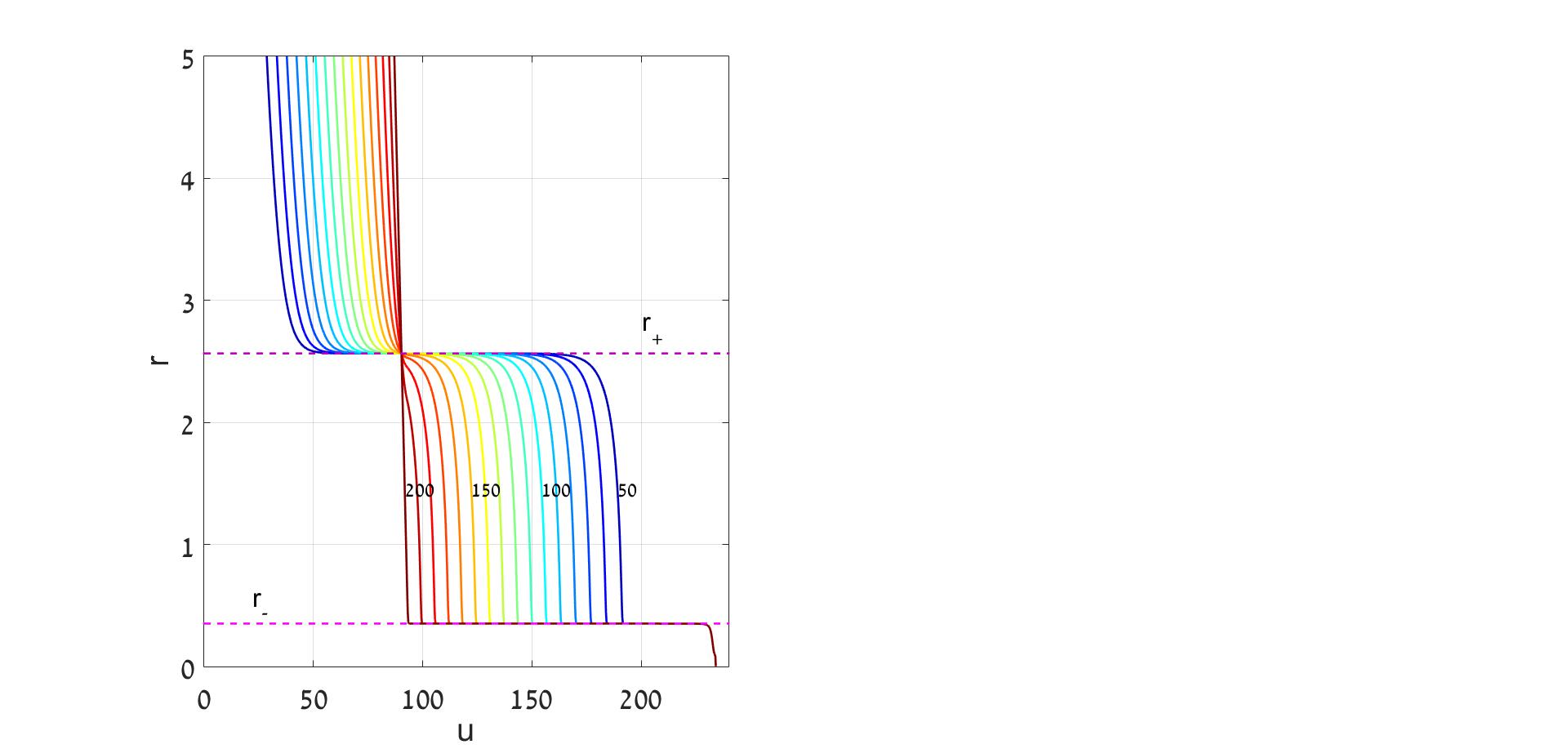}
\par\end{centering}

\protect\caption{\label{fig:r_of_u_RN-SF} $r$ as a function of $u$ along various
late $v=const$ grid rays, in the presence of self-gravitating scalar
field. Shown are sixteen $v=const$ rays in the range $50\leq v\leq200$
(with fixed increments of $10$), in different colors, with $v=50$
in blue and $v=200$ in dark red. The graph actually includes two
resolutions for each grid ray, $N=320$ and $N=640$; but the two
resolutions overlap. The horizontal purple dashed line represents
$r_{+}=2.5655$; the horizontal magenta dashed line represents $r_{-}=0.3517$.
The graph demonstrates some aspects of the anticipated asymptotically-RN
spacetime: (i) all curves cross at the event horizon (at $u\approx90.3$),
and (ii) as $u$ increases, all curves join a ``universal curve'',
which is approximately located at $r=r_{-}$ (except at very large
$u$). The main feature which differs from RN is the sharp fall of
this universal curve toward $r=0$ at around $u\sim230$. This sharp
fall results from the focusing effect of the scalar field.}
\end{figure}

\begin{figure}[H]
\noindent \begin{centering}
\includegraphics[bb=120bp 0bp 1760bp 944bp,clip,scale=0.3]{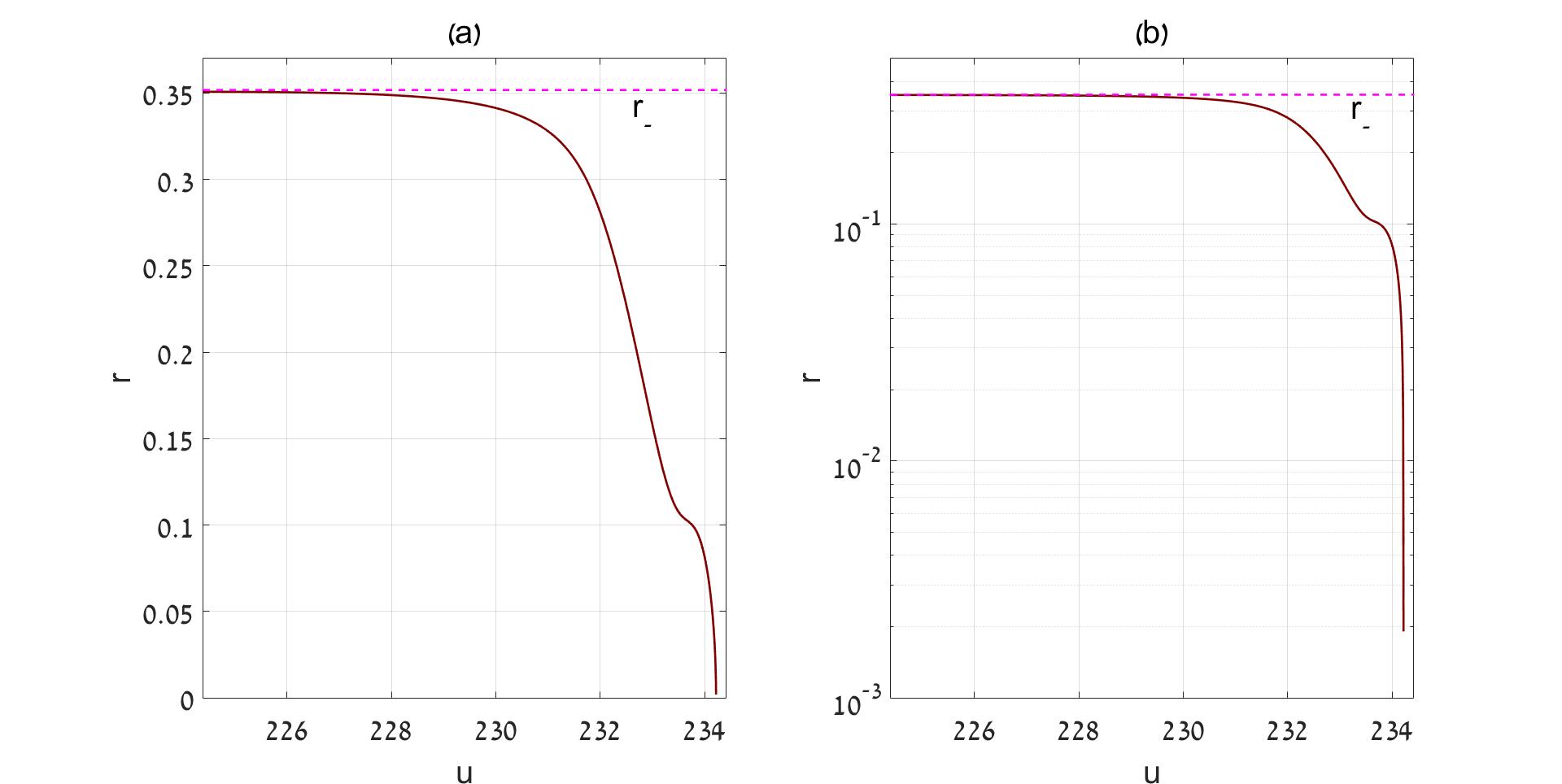}
\par\end{centering}

\protect\caption{\label{fig:r_of_u_RN-SF-zoom} The contraction of Cauchy Horizon as
reflected in the function $r(u)$ along various late $v=const$ rays.
This figure is similar to figure \ref{fig:r_of_u_RN-SF}, but zoomed
on the final drop of $r$ toward the \emph{$r=0$ }singularity. Panel
(a) uses a linear scale for $r$, and panel (b) uses a logarithmic
scale. Notice that all the $v=const$ rays (in the range $50\leq v\leq200$
shown here) overlap on this scale, as well as the two resolutions
$N=320$ and $N=640$ for each ray.}
\end{figure}

Figure \ref{fig:r_of_u_RN-SF} displays $r(u)$ along rays of constant
$v$. This figure demonstrates two interesting (although well known)
facts: First, while $u$ increases the various constant-$v$ curves
join a universal limiting curve (the red curve at the bottom), which
should be interpreted as $r(u)$ along the CH. Second, the CH undergoes
a clear contraction at large $u$: The CH function $r(u)$ is initially
``frozen'' at $r\approx r_{-}$, but at large $u$ it starts shrinking
toward $r=0$. This behavior is demonstrated more clearly in Fig.
\ref{fig:r_of_u_RN-SF-zoom} which zooms on the large-$u$ region
in Fig. \ref{fig:r_of_u_RN-SF}. This contraction of the CH to $r=0$
results from the focusing effect due to the scalar-field energy fluxes.
Panel (b) provides a logarithmic graph of this function $r(u)$, indicating
that the numerics performs pretty well up to $r$ values as small
as $\sim2\cdot10^{-3}$.

\noindent 

\noindent 

\begin{figure}[H]
\noindent \begin{centering}
\includegraphics[bb=130bp 0bp 1800bp 944bp,clip,scale=0.3]{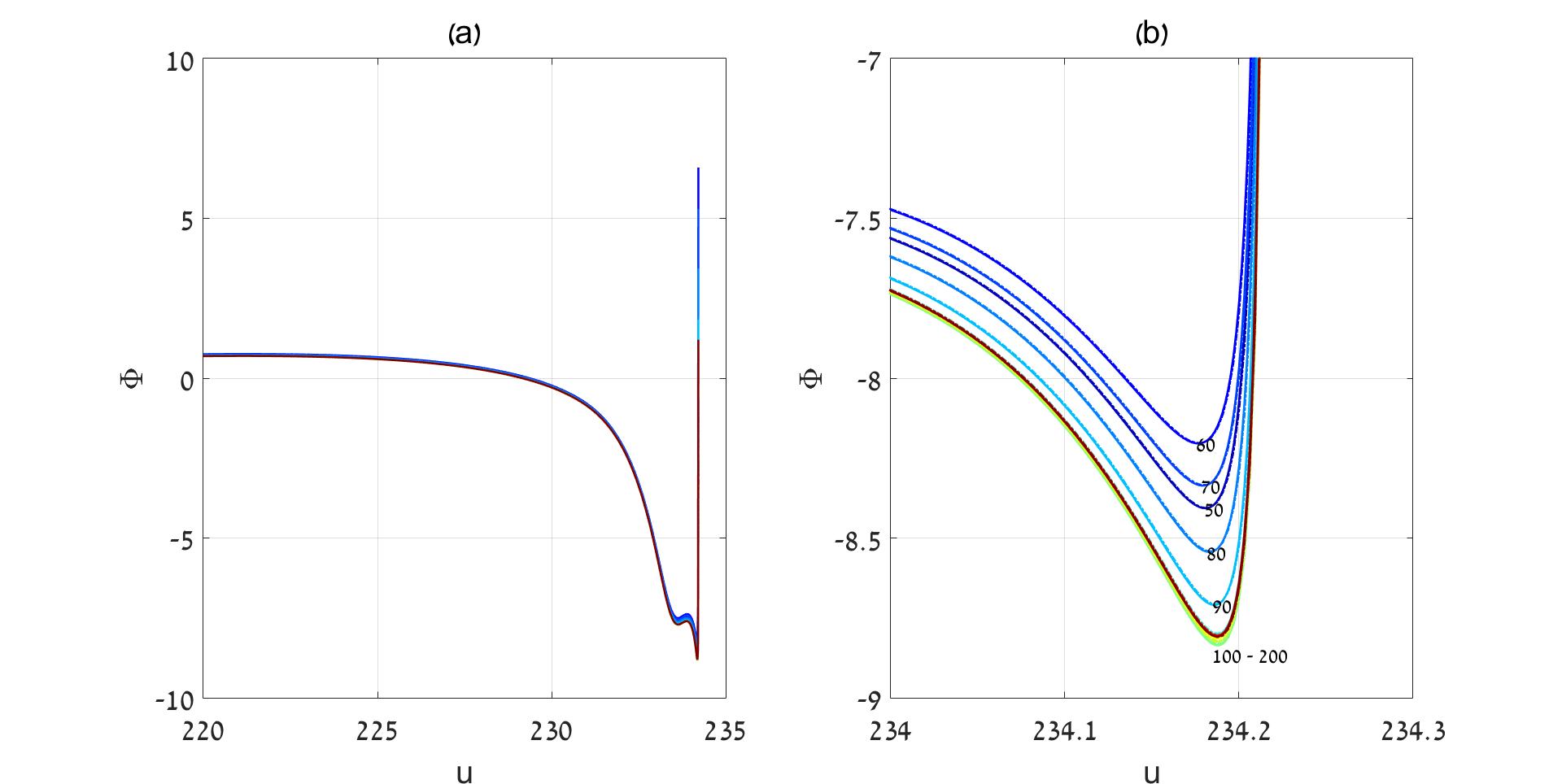}
\par\end{centering}

\protect\caption{\label{fig:Phi_of_u_RN-SF-1} $\Phi$ as a function of $u$ along
various late $v=const$ grid rays, in the presence of self-gravitating
scalar field. Here too the $v=const$ rays are in the range $50\leq v\leq200$.
Panel (b) is a zoom of panel (a). Both panels include two resolutions
for each grid ray, $N=320$ (dashed curves) and $N=640$ (solid curves);
the two resolutions overlap in the panel (a), but a minor deviation
between the solid and dashed curves is barely seen in panel (b). The
numbers in panel (b) indicate the $v$ values of the various curves.
Notice the non-monotonic ordering of these numbers, which indicates
oscillations of $\Phi$ with $v$ (at fixed $u$), in the range $50\leq v\lesssim100$.
However, at larger $v$ values ($100\lesssim v\leq200$) the oscillations
are replaced by a monotonic convergence toward the CH.}
\end{figure}

\begin{figure}[H]
\noindent \begin{raggedright}
\includegraphics[bb=100bp 0bp 1820bp 944bp,clip,scale=0.3]{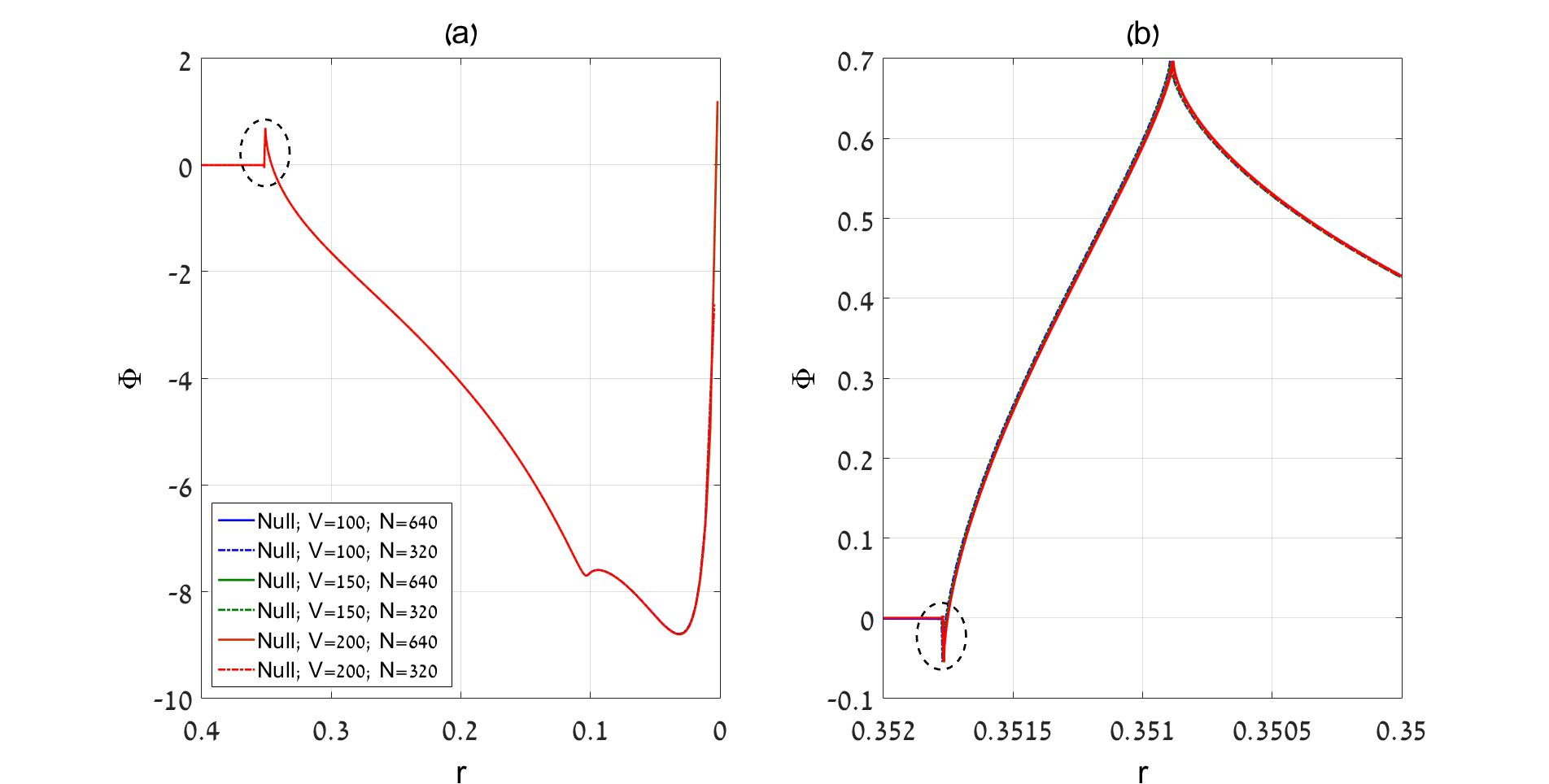}
\par\end{raggedright}

\noindent \begin{raggedright}
\includegraphics[bb=100bp 0bp 1820bp 944bp,clip,scale=0.3]{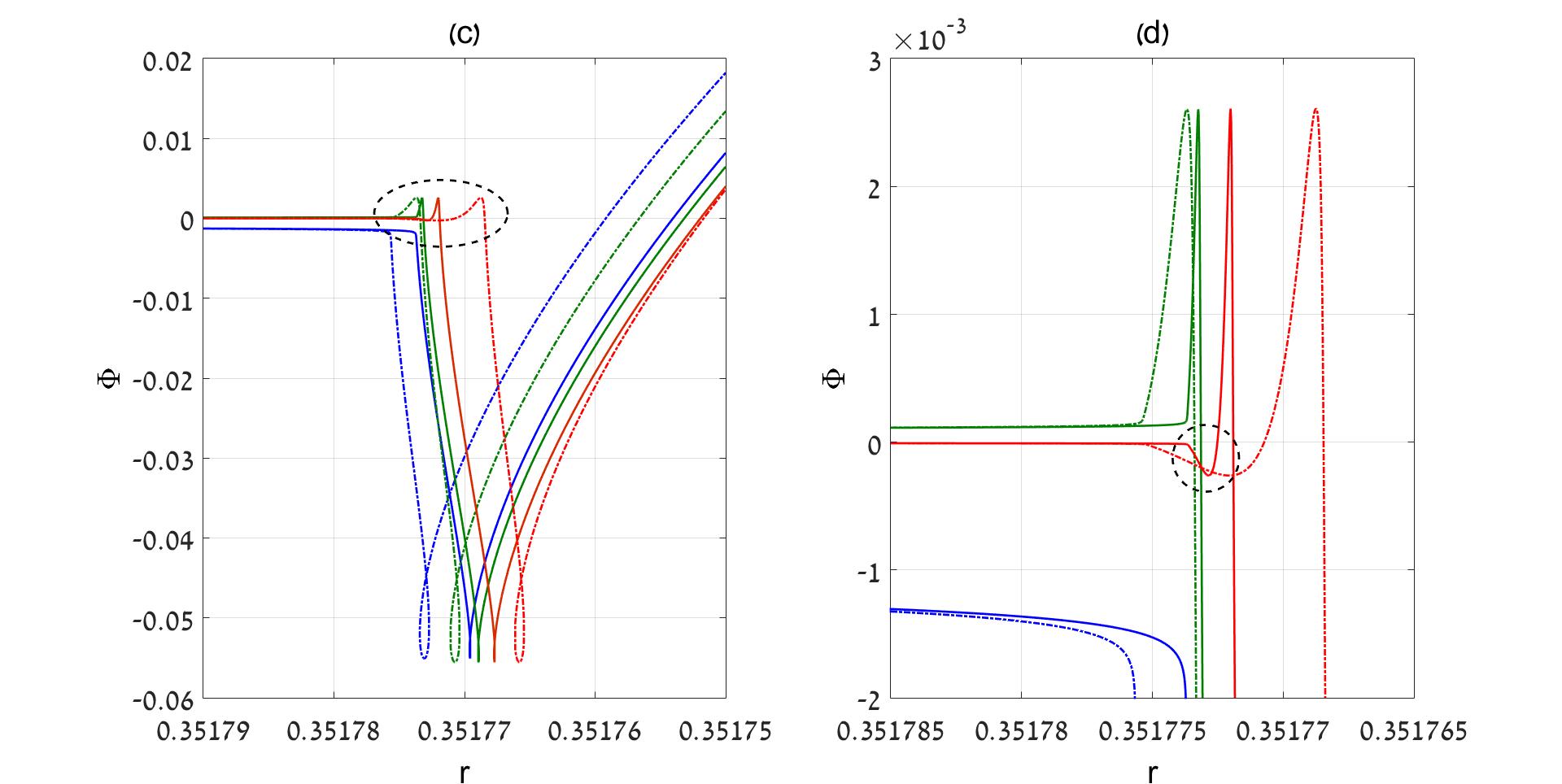}
\par\end{raggedright}

\protect\caption{\label{fig:Phi_of_r} The dependence of the scalar field $\Phi$ on
$r$ along various late $v=const$ grid rays deep inside the charged
black hole. All panels contain two resolutions for each curve ($N=320$
and $N=640$). Panel (a) displays the zone $0\leq r\lesssim r_{-}$.
Notice the spike (circled), at around $r\sim0.35$. Panels (b),(c)
and (d) zoom on this spike, each of them revealing a new spike. The
spike in panels (c,d) are located fairly close to the anticipated
inner-horizon value $r\approx0.3517$. The spike in (c) is evident
only in the late rays $v=150$ and $v=200$. (The \textramshorns -like
shape of the curves of lower resolution $N=320$ is a numerical artifact.)
The spike in panel (d) is only evident in the last ray $v=200$, and
only in the best resolution $N=640$. The overall picture emerging
from these panels suggests the existence of hierarchical structure,
in which each feature is followed by another, even sharper one, of
opposite sign and smaller amplitude.}
\end{figure}

Figure \ref{fig:Phi_of_u_RN-SF-1} displays $\Phi(u)$ along various
curves of constant $v$. The sharp increase around $u\approx234$
seems to follows from the very significant $r$-focusing, which occurs
at that $u$ value as can be seen in panel (b) of Fig. \ref{fig:r_of_u_RN-SF-zoom}.
A zoom on the sharp (negative) peak at $u\approx234.2$ is shown in
panel (b). One can notice the convergence of the various $v=const$
curves towards the red one ($v=200$) as $v$ increases. This represents
the decay of the scattering dynamics at large $v$ (``issue (ii)''
in the discussion above). 

\noindent 

Figure \ref{fig:Phi_of_r} displays $\Phi$ as a function of $r$,
again along various constant-$v$ rays. In some sense this function
$\Phi(r)$ is more robust than $r(u)$ and $\Phi(u)$, because it
is invariant to the $u$-gauge. And furthermore, its large-$v$ limit
(namely the CH limiting function), which we may denote $\Phi_{ch}(r)$,
is entirely gauge-invariant. 

A strange spike shows up in panel (a) at $r\sim0.35$ (circled). A
zoom on this spike, shown in panel (b), successfully resolves the
steep increase in $\Phi$. However, it reveals a new spike (this time
a negative one) at $r\approx0.3518$. The further zoom in panel (c)
reveals yet another spike. This situation repeats itself once again
in the next zoom level shown in panel (d), indicating that this feature
actually enfolds a \emph{hierarchy of spikes}, with alternating signs.
The spikes in panel (c,d) are located at $r\approx0.351773,$ which
agrees fairly well with the $r_{-}$ value corresponding to the final
mass $m_{final}=1.459$. The origin and properties of this series
of spikes remains as an open question for further investigation.

\noindent 

\section{Discussion\label{sec:Discussion}}

As was shown in the previous sections, our algorithm solves the difficulties
associated with long-run double-null simulations along the event horizon
of a spherical BH. It also solves the analogous problem that occurs
deeper inside a charged BH, at the outgoing inner horizon. It thus
allows a full and continuous simulation of the spacetime of a spherical
charged BH: from the weak-field region, throughout the strong-field
region and across the EH; and in the BH interior, up to the neighborhoods
of the CH and the $r=0$ singularity (as well as their intersection
point).

There still remain the usual limitations of the finite-difference
approach, namely the accumulation of the truncation and roundoff errors
--- which may eventually limit the integration range $\delta v_{e}$.
Also, it seems that it would not be easy to parallelize this algorithm,
due to the non-trivial set-up of the initial data for $\sigma$. 

Our main motivation for developing this algorithm was the attempt
to numerically explore the outgoing shock wave that was recently found
\cite{2012-Marolf-Ori_Shockwave} to occur at the outgoing inner horizon
of a generically-perturbed spinning or charged BH. This numerical
investigation of the shock wave will be presented elsewhere. However,
we believe that the same adaptive-gauge algorithm may be used for
a variety of other investigations of spherically-symmetric self-gravitating
perturbations which require a long-term simulation (namely large span
in Eddington $v$) along the event and/or outgoing inner horizon.
For example it could be used to investigate the evolution of ingoing
power-law tails of non-linear perturbations inside the BH, and particularly
near the CH. 

It has been argued by Hamilton and Avelino \cite{2010-Hamilton-Avelino}
that the description of the CH as a null singularity \cite{1981-Hiscock-PRN-Analytical}\cite{1990-Poisson-Israel-PRN-Analytical-mass-function}\cite{1991-Ori-PRN-Analytical}\cite{1992-Ori-spinning}
is not relevant to astrophysical BHs since they are not isolated.
Instead, realistic BHs steadily accrete the microwave background photons,
and possibly also dust. Hamilton and Avelino argued that these long-term
fluxes should drastically change the nature of the null weak singularity
at the (would-be) CH, replacing it by a stronger singularity. Our
algorithm can be used to clarify this issue, by numerically exploring
the interior of a spherical charged BH perturbed by null fluids and/or
self-gravitating scalar field, which fall into the BH along a very
long period of Eddington $v$ --- e.g. of order $10^{3}M$. Although
the code presented in this paper only included a scalar field, the
addition of null fluids to the code is fairly straightforward, it
merely involves a minor modification in the constraint equations. 

Finally, it is worth mentioning that while our simulation is fully
classical, the double-null EH problem is not restricted to classical
perturbations and is also relevant to semi-classical dynamics. In
particular, in any attempt to simulate the evolution of evaporating
BH due to weak semi-classical fluxes, a significant decrease in the
BH mass will require a very long evolution time. The aforementioned
EH problem will then be encountered if double-null coordinates are
used. Our adaptive gauge algorithm may be employed to overcome the
EH problem in this case as well.

\section*{Acknowledgments}

We are grateful to Dong-han Yeom for bringing Refs. \cite{2003-Oren-Piran - Adaptive mesh,2005-Hansen Khokhlov-Novikov - Adaptive mesh,2010-Hong-Yeom et. al - Adaptive Mesh}
to our attention. This research was supported by the Israel Science
Foundation (Grant No. 1346/07).

\appendix

\section{Matter fields and their energy-momentum tensor\label{sec: Appendix-A}}

\noindent Our model contains two matter fields: A minimally-coupled
massless scalar field $\Phi$, and a radial electromagnetic field
$F_{\mu\nu}$. We shall explicitly express here the field equations
and also the form of the Energy-Momentum tensor for these two fields.

\subsection{Scalar field}

The field equation for a minimally-coupled massless scalar field is
$\Phi_{;\alpha}^{\,\,;\alpha}=0$. In our double-null coordinates
it takes the explicit form
\begin{equation}
\Phi,_{uv}=-\frac{1}{r}(r,_{u}\Phi,_{v}+r,_{v}\Phi,_{u}).\label{eq:  phi_evolutio-1}
\end{equation}
The Energy-Momentum tensor of this field is

\begin{equation}
T_{,\mu\nu}^{\Phi}=\frac{1}{4\pi}(\Phi,_{\mu}\Phi,_{\nu}-\frac{1}{2}g_{\mu\nu}g^{\alpha\beta}\Phi,_{\alpha}\Phi,_{\beta}).
\end{equation}
With the line element (\ref{eq:line-element}) this reads 
\begin{equation}
T_{\mu\nu}^{\Phi}=\frac{1}{4\pi}\left(\begin{array}{cccc}
\Phi,_{u}^{2} & 0 & 0 & 0\\
0 & \Phi,_{v}^{2} & 0 & 0\\
0 & 0 & 2r^{2}\,\Phi,_{u}\Phi,_{v}e^{-\sigma} & 0\\
0 & 0 & 0 & 2r^{2}\sin^{2}\theta\,\Phi,_{u}\Phi,_{v}e^{-\sigma}
\end{array}\right)\,.\label{eq:Energy-Momentum-scalar}
\end{equation}

\subsection{Electromagnetic field}

The electromagnetic field $F_{\mu\nu}$ satisfies the source-free
Maxwell equation. Restricting attention to spherically-symmetric solutions
which entails a radial electric field, it is fairly straightforward
to solve these equations for a generic (time-dependent) spherically-symmetric
background metric as in Eq. (\ref{eq:line-element}). The source-free
Maxwell equation is given by:

\begin{equation}
F^{\mu\nu}\thinspace_{;\nu}\equiv\frac{1}{\sqrt{-g}}(\sqrt{-g}F^{\mu\nu}),{}_{\nu}=0,\label{eq:Free-Maxwell-Eq}
\end{equation}

\noindent where $g=-(1/4)e^{2\sigma}r^{4}\sin^{2}\theta$ is the metric
determinant. In our case (radial electric field) the only non-vanishing
component of $F^{\mu\nu}$ is $F^{uv}=-F^{vu}$. Correspondingly $F^{uv}$
satisfies 

\begin{equation}
(\frac{1}{2}e^{\sigma}r^{2}F^{uv}),{}_{v}=0\,,\,(\frac{1}{2}e^{\sigma}r^{2}F^{uv}),{}_{u}=0
\end{equation}

\noindent and integration yields:

\begin{equation}
\frac{1}{2}e^{\sigma}r^{2}F^{uv}=const\equiv Q.
\end{equation}

\noindent Here $Q$ is a free parameter that should be interpreted
as the field's electric charge. This equation yields $F^{uv}=2Qe^{-\sigma}/r^{2}$;
using $g_{uv}=-e^{\sigma}/2$ in order to lower indices we obtain
the final result for the covariant components:

\begin{equation}
F_{vu}=-F_{uv}=Q\,\frac{e^{\sigma}}{2r^{2}}.\label{eq:EM-solution}
\end{equation}

The energy-Momentum tensor of the electromagnetic field is given by
\[
T_{\mu\nu}^{Q}=\frac{1}{4\pi}(g^{\alpha\beta}F_{\mu\alpha}F_{\nu\beta}-\frac{1}{4}g_{\mu\nu}F_{\alpha\beta}F^{\alpha\beta}).
\]
In our coordinates this reads

\begin{equation}
T_{\mu\nu}^{Q}=\frac{Q^{2}}{8\pi r^{4}}\left(\begin{array}{cccc}
0 & \frac{1}{2}e^{\sigma} & 0 & 0\\
\frac{1}{2}e^{\sigma} & 0 & 0 & 0\\
0 & 0 & r^{2} & 0\\
0 & 0 & 0 & r^{2}\sin^{2}\theta
\end{array}\right).\label{eq:Energy-Momentum-EM}
\end{equation}

\end{document}